\documentclass[a4paper,11pt]{article}
\usepackage[margin=1in]{geometry}
\usepackage{authblk}
\date{}
\usepackage{amsmath,amssymb,amsfonts}
\usepackage{graphicx}
\usepackage{multicol,multirow}
\usepackage{mathrsfs}
\usepackage{amsthm}
\usepackage{rotating}
\usepackage{booktabs,multirow,dcolumn,adjustbox,threeparttable,siunitx,float,graphics,breakcites}
\usepackage{rotating} 
\usepackage{pdflscape}
\usepackage{color}
\usepackage[round]{natbib} 
\usepackage{caption}
\usepackage{subcaption}
\usepackage{hyperref}
\usepackage{etoolbox}
\usepackage{placeins}
\usepackage{algorithm}
\usepackage{algorithmic}
\usepackage[english]{babel}
\AtBeginEnvironment{sidewaystable}{}
\newcolumntype{d}[1]{D{.}{.}{#1}}  
\newcommand\englishkeywordslabel{Keywords:}
\newcommand\englishkeywords[1]{%
  \begin{list}{}{%
    \setlength{\topsep}{2ex}%
    \settowidth{\leftmargin}{\bfseries\englishkeywordslabel~}%
    \setlength{\labelsep}{0pt}%
    \setlength{\labelwidth}{\leftmargin}%
    \setlength{\itemindent}{0pt}%
  } 
  \raggedright\item[\englishkeywordslabel~]#1
  \end{list}
}
\newtheorem{theorem}{Theorem}[section]

\theoremstyle{definition}
\newtheorem{definition}[theorem]{Definition}
\newtheorem{remark}[theorem]{Remark}

\numberwithin{equation}{section}

\title{Forecasting mortality rates with functional signatures}
\author[1]{Yap Zhong Jing}
\author[1,2,3]{Dharini Pathmanathan}
\author[4,5]{Sophie Dabo-Niang}
\affil[1]{Institute of Mathematical Sciences, Faculty of
Science, Universiti Malaya, 50603 Kuala Lumpur, Malaysia}
\affil[2]{Universiti Malaya Centre for Data Analytics, Universiti Malaya, 50603 Kuala Lumpur, Malaysia}
\affil[3]{Center of Research for Statistical Modelling and Methodology, Faculty of Science, Universiti Malaya, 50603 Kuala Lumpur, Malaysia}
\affil[4]{CNRS, UMR 8524-Laboratoire Paul Painlevé,
INRIA-MODAL, Université Lille, F-59000 Lille,
France}
\affil[5]{CNRS—Université de Montréal, CRM—CNRS, Montréal, Canada}

\begin{document}
\maketitle
\begin{abstract}
This study introduces an innovative methodology for mortality forecasting, which integrates signature-based methods within the functional data framework of the Hyndman-Ullah (HU) model. This new approach, termed the Hyndman-Ullah with truncated signatures (HUts) model, aims to enhance the accuracy and robustness of mortality predictions. By utilizing signature regression, the HUts model is able to capture complex, nonlinear dependencies in mortality data which enhances forecasting accuracy across various demographic conditions. The model is applied to mortality data from 12 countries, comparing its forecasting performance against variants of the HU models across multiple forecast horizons. Our findings indicate that overall the HUts model not only provides more precise point forecasts but also shows robustness against data irregularities, such as those observed in countries with historical outliers. The integration of signature-based methods enables the HUts model to capture complex patterns in mortality data, making it a powerful tool for actuaries and demographers. Prediction intervals are also constructed with bootstrapping methods.
\end{abstract}
\englishkeywords{Hyndman-Ullah model, Lee-Carter model, principal component analysis, functional data analysis}

\section[Introduction]{Introduction}

Mortality modeling and forecasting are quintessential to the disciplines of actuarial science, demography, and public health, serving as critical tools for socio-economic planning and risk management. Their primary objective is to provide reliable estimates of future mortality rates and life expectancies, which are essential for shaping policy and financial strategies. For government agencies, anticipating mortality trends is fundamental for effective resource allocation in healthcare, social security, and public welfare sectors. Precise mortality forecasts enable informed decisions on retirement ages, healthcare funding, and pensions, ensuring that social security systems remain robust amid demographic changes.

In the insurance and pension industries, mortality modeling takes part in the pricing of life insurance, annuities and managing pension funds. Actuaries rely on these models to calculate premiums, reserves, and pension contributions that align with predicted mortality rates, thus safeguarding institutions against longevity risks \citep{Deprez2017}. Accurate mortality projections are also critical in addressing global challenges posed by increasing life expectancy and declining fertility rates which shift the age structure of populations. These demographic trends have profound implications for the economic sustainability of pension systems and public health policies \citep{Bjerre2022}. Consequently, the practice of mortality modeling and forecasting not only supports financial and actuarial operations but also facilitates broader strategic planning and policy-making to accommodate the evolving demographic landscapes, ultimately ensuring the economic and social stability of societies.

\subsection{Mortality modelling and forecasting}
Mortality models are broadly categorized into three approaches: expectation, explanation, and extrapolation \citep{Booth_Tickle_2008}. Each serves distinct purposes in understanding and predicting mortality rates. The expectation approach relies on expert opinions to forecast mortality, emphasizing informed conjectures based on past experiences. Meanwhile, explanation models link mortality rates to various risk factors such as lifestyle or socioeconomic status, using regression techniques to uncover the underlying causes of mortality patterns. The primary focus, however, is on extrapolation, the most actively researched and commonly used approach, particularly by national statistical bureaus. This method projects future mortality rates based on the continuation of observed historical trends and age patterns. By leveraging time series analysis and other statistical techniques that emphasize trend continuity, extrapolative models provide robust forecasts grounded in the regularity of past data, making them indispensable for demographic analysis and policy planning.

The seminal work of \citet{LeeCarter1992} was undoubtedly one of the most influential extrapolative models to be introduced to forecast mortality rates. While the Lee-Carter (LC) model was initially applauded for its simple structure and straightforward implementation, it did have its fair share of limitations which were subsequently addressed in various extensions of the LC model \citep{LeeMiller2001,Booth2002}. Extensions were also not limited to adjustments to the original LC model as seen in the deployment of generalised linear models \citep{RenshawHaberman2003} and Bayesian statistics \citep{Czado2005}. Readers are directed to \citet{BASELLINI2023} for a more recent review of LC methods since the introduction of the LC model. 

The LC extension of interest in this paper involves the functional data paradigm \citep{RamsaySilverman2005}. The Hyndman-Ullah (HU) model \citep{HYNDMAN20074942} is a generalisation of the LC model that uses concepts from functional data analysis to better model mortality rates by exploring variations within the data from a functional perspective. The use of nonparametric smoothing techniques on the age-specific mortality rates allowed for the removal of the LC model's homoskedasticity assumption and realise an underlying smooth curve that can be decomposed using functional principal component analysis with their orthonormal basis. Component wise it is similar to the LC model except it allows for multiple principal components to be included, accounting more variation into the model which \citet{HYNDMAN20074942} noted is the reason for its superior performance for longer-term forecast. Comparisons also show that the HU model outperformed the LC, Lee-Miller, and Booth-Maindonald-Smith models. Recent extensions of the HU model have been for the modelling of subpopulations \citep{cristian2024,SHANG2022239,shang2017}. In light of the substantial contributions made by the HU model, there is still a growing interest to explore new modifications to improve its efficacy and adaptability. 

With the rise of network approaches in modelling, there is growing interest in extending age-period models using deep learning techniques. Various aspects of the LC model have been coupled with deep learning applications, both in parameter estimation \citep{Schnürch_Korn_2022,Scognamiglio_2022,Richman_Wüthrich_2021,Hainaut_2018,Miyata_Matsuyama_2022} and in forecasting procedures \citep{perla2021,Bjerre2022,nigri,marino2023}. Deep learning models fundamentally process streams of input data to approximate relationships. While they perform well, they can lack replicability and often require increased computational resources. Signatures offer a potential solution to these limitations as they too possess universality to approximate relationships arbitrarily well, reproducible, and efficient. Therefore, this paper aims to integrate signatures that better capture the nonlinear patterns often present in single-population mortality data into the HU model's functional data framework.

\subsection{Signatures from rough path theory}
The signature is a sophisticated mathematical tool that is capable of summarising complex, high-dimensional data paths through their iterated integrals in an informative and compact package as a power series that contains the tensor coefficients. Originally introduced by \citet{Chen1957} and further developed in the context of rough path theory by \citet{Lyons2007}, they have found extensive applications across multiple fields, particularly where the analysis of dynamic systems is crucial. Their effectiveness has been demonstrated in various applications such as sepsis detection \citep{MorrillSepsis}, identifying Chinese handwriting \citep{Xie2018}, and in diagnosing psychological disorders \citep{wang20e_interspeech}. The increasing interest in signatures stems primarily from their ability to effectively condense information from complex sequences without losing critical details, making them exceptionally useful for detection work. This efficiency in capturing the essence of data paths while minimizing complexity is what drives their growing popularity, especially in areas dealing with large volumes of sequential data such as in the field of quantitative finance \citep{lyons2020,jasdeep,cuchiero2024signaturemethodsstochasticportfolio,jaber2024signaturevolatilitymodelspricing}.

Signatures possess a unique mathematical property known as universal nonlinearity, which enables the application of regression techniques directly on these signature representations, known as signature regression. This method leverages the rich algebra provided by the signatures to model and predict dependent variables from a sequence of inputs. The practical implication of this is that it can construct predictive models that are capable of handling data with complex temporal dynamics and interactions. In other words, the universality of signatures means that any continuous function on the space of signatures can approximate continuous functions on the space of paths to an arbitrary degree of accuracy. This broad capability makes signature-based models exceptionally versatile and powerful, providing a robust framework for tackling problems where conventional regression models might struggle with the complexities of the data.

Signature regression is further highlighted through various applications across different fields. For instance, \citet{levin2016} utilized the universal properties of signatures to characterize functional relationships within datasets, leading to the development of the expected signature model. This model leverages linear regression on signature features to summarize the conditional distribution of responses effectively. It has demonstrated superior forecasting abilities compared to autoregressive and Gaussian Process models, especially in scenarios involving large datasets, thereby highlighting its efficiency in data stream summarization and significant dimension reduction. 

Moreover, the adaptability and simplicity of modelling with signature regression have been elaborated by \citet{cohen2023nowcasting}, who noted their effectiveness in simplifying time series modeling. This approach involves merely computing signatures from observed paths and applying regression, significantly reducing the complexity associated with time series analysis. The flexibility of signature methods also extends to handling missing data and irregular sampling, making them applicable for diverse forecasting and analysis tasks. Notably, when applied to nowcast the gross domestic product growth of the United States, signature regression methods have outperformed dynamic factor models, showcasing its potential in economic forecasting.

Building on these insights, \citet{FERMANIAN2022} was the first to apply the concept of signatures to functional data analysis by introducing the signature linear model within a functional regression framework. \citet{FERMANIAN2022} noted that signatures are naturally adapted to vector-valued functions, unlike other functional methods, and require little assumption on the regularity of functions to encode nonlinear geometric information. This model operates by mapping functions to their signatures and employing ridge regression for prediction, showing enhanced capabilities in handling multidimensional data. By surpassing established methods like functional linear models and functional principal component regression, the signature linear model eliminates the need for certain assumptions typically required within functional settings, streamlining the modeling process. This lays the foundation that suggests that signature regression is a good candidate to extend the functional data framework of the HU model.

In the field of actuarial science, the use of signatures has started to make inroads, particularly in areas such as validating economic scenarios \citep{Andrès_Boumezoued_Jourdain_2024}, and option pricing \citep{cuchiero_siam,cuchiero2024jointcalibrationspxvix}. However, the application of signatures in the life and mortality side of actuarial science remains less explored, indicating a gap for research and potential development. The introduction of signature regression into mortality forecasting presents a novel methodology that could improve the accuracy of predictions by leveraging the universality of signatures to effectively model complex, nonlinear relationships inherent in mortality data. It is therefore the aim of this article to introduce a functional mortality model by which is decomposed using truncated signatures in place of FPCA that also utilises the extrapolation method to produce forecasts. The model will from here on be referred to as the HU with truncated signatures (HUts) model and will be used to obtain point and interval forecasts of log mortality rates of several countries that are readily available from the \citet{HMD}. 

The remainder of this paper is structured as follows. Section 2 provides a description of the methodology, detailing the HU model and the integration of signature regression. Section 3 presents the empirical analysis, including data preparation and results of applying the proposed model to mortality data across various countries. In Section 4, we discuss the implications of these results, and section 5 concludes.

\section{Methodology}

\subsection{The Hyndman-Ullah (HU) model}

The HU model is a generalised extension of the LC model, employing the functional data paradigm \citep{RamsaySilverman2005} and nonparametric smoothing to improve the forecasting of mortality rates. This model is distinguished by its use of multiple principal components, capturing a broader variation within the data, and applying more advanced time series forecasting methods than the random walk with drift model typically used in the LC model \citep{HYNDMAN20074942}.

At its core, the HU model assumes that log mortality rates $y_t(x)$ observed at discrete ages $x$ and years $t$ have an underlying smooth function \( f_t(x) \), observed with error:

\begin{equation}
    y_t(x) = f_t(x) + \sigma_t(x)\epsilon_t, \nonumber
\end{equation}

\noindent where $\epsilon_t$ is the standard normal error term and $\sigma_t(x)$ denotes the observational standard deviation which is allowed to vary with age. This lets the HU model handle data heterogeneity and ensures demographic consistency by applying monocity constraints when using penalized regression splines to smooth the mortality curves \citep{HYNDMAN20074942}.

Functional principal component analysis (FPCA) is used to decompose the smoothed mortality rates into principal components. By applying FPCA to the centered smooth mortality would yield:

\begin{equation}
    f_t(x) = \mu(x) + \sum_{k=1}^K \beta_{t,k} \phi_k(x) + e_t(x). \nonumber
\end{equation}

\noindent Here, $\mu(x)$ represents the mean function, $\phi_k(x)$ are orthonormal basis functions, $\beta_{t,k}$ are time dependent coefficients, and $e_t(x)$ is the error function.  Forecasts of mortality rates are done through the time dependent coefficients, $\beta_{t,k}$ which are fitted into univariate ARIMA models and extrapolated $h$-steps-ahead.

A critical consideration in applying FPCA is its sensitivity to outliers. Extreme values can disproportionately influence the principal components derived from the data. This sensitivity may lead to biased or misleading results, especially when modeling phenomena such as mortality rates where spikes might occur due to extraordinary events like epidemics or wars. To address these issues, various adaptations of the HU model have been put forth. These include the robust HU (HUrob) model with robust FPCA \citep{HYNDMAN20074942}, which seeks to minimize the impact of outliers by using the $L_1$-median as the location measure, paired with the reflection-based algorithm for principal components analysis \citep{HUBERT2002101} for a robust set of principal components; the weighted HU (wHU) model with weighted FPCA \citep{HYNDMAN2009}, which adjusts the influence of data points by placing more weight on recent data with the measure of location being a weighted average using geometrically decreasing weights. These modifications aim to enhance the stability and reliability of the model under conditions where data anomalies are present. Having the same objective in mind to make the model less sensitive to outliers, we extend the literature by exploring the use of signature regression on the signature of age specific mortality rates as an alternative to enhancing the robustness of mortality forecasts.

\subsection{The signature of a path}
\subsubsection{The concept of signatures}
In statistical analysis, particularly when dealing with time-series or sequential data, understanding the concept of a ``path'' is crucial. A path $X: [0,1] \rightarrow \mathbb{R}^d$ is defined as a continuous mapping from an interval to a topological space, which is assumed to be piecewise differentiable and of bounded variation. Unlike time series or stochastic processes, which are explicitly governed by time, a path emphasizes its geometrical movements through space, making the term "path" more suitable. From a computational perspective, think of a path as input data with $d$ dimensions, analogous to multi-dimensional time series data.

Signatures are mathematical tools used to encode the information contained in a path. Using Chen's identity \citep{Chen1957}, the signature of a piecewise linear path is computed by taking the tensor product of exponentials of the linear segments of the path. Thus, the signature of a path, or simply signatures, is a collection of all iterated integrals between coordinates of a path. This collection is often described as an abstract summary of the input signal's variability. 

\begin{definition}[Signature of a path \citep{FERMANIAN2022}]
    \label{DefSig}
  \textit{Let $X:[0,1]\rightarrow \mathbb{R}^d$ be a path of bounded variation, and let $I=(i_1,\ldots,i_k)\in\{1,\ldots,d\}^k$, where $k\in\mathbb{N}^*$, be a multi-index of length $k$. The signature coefficient of $X$ corresponding to the index $I$ is defined by}
  \begin{align}
    \label{SigCo}
    S^I(X) & = \idotsint\limits_{0 \leq u_1 < \dots < u_k \leq 1} dX_{u_1}^{i_1} \dots dX_{u_k}^{i_k} \nonumber \\
    & = \int_{0}^{1}\int_{u_1}^{1} \int_{u_2}^{1} \dots \int_{u_{k-1}}^{1} dX_{u_k}^{i_k} \dots dX_{u_2}^{i_2} dX_{u_1}^{i_1}\text{,} \notag
  \end{align}
\textit{where $S^I(X)$ is called the signature coefficient of order $k$. The signature of $X$ is then the sequence containing all the signature coefficients:}
 \begin{equation}
    \label{SigPath}
    S(X)=(1, S^{(1)}(X), \ldots, S^{(d)}(X), S^{(1,1)}(X), \ldots, S^{(i_1,\ldots,i_k)}(X), \ldots). \nonumber  
 \end{equation}
\end{definition}

However, from a practical standpoint, it is not feasible to use all the signature coefficients in the infinite sequence of the signature of $X$ due to computational limits, and also due to the factorial decay of the coefficients as the order increases \citep{bonnier2019deep}. Therefore, it is often sufficient to consider the signature of a path up to a truncated order of $m$, denoted as:
 \begin{equation}
    \label{SigPathTrunc}
    S^m(X)=(1, S^{(1)}(X), \ldots, S^{(d)}(X), S^{(1,1)}(X), \ldots, S^{(i_1,\ldots,i_m)}(X)), \nonumber
\end{equation}

\noindent which contains $s_d(m)=\frac{d^{m+1}-1}{d-1}$ number of signature coefficients. Note that the number of signature coefficients grows exponentially with the truncation order.

A fundamental attribute of the signature is its ability to encode the geometric properties of a path. The first-order terms, $S^{(i)}(X)$, are straightforward, representing the total change along each dimension of the path or the  increments observed in each variable. The second-order terms, $S^{(i_1, i_2)}(X)$, provide insights into the interaction between pairs of dimensions, similar to capturing the area enclosed by the path as it travels through these dimensions, and higher-order interactions are particularly valuable for understanding complex dependencies and dynamics within the data.

\subsection{Embedding with signatures}
Embedding paths before computing its signature is a standard practice used to maximise the potential of signatures by addressing inherent invariances such as translation and time reparametrization that can result in significant information loss. These embedding transformation ensure that all relevant information are preserved and encoded into the signature and in turn enrich the signature’s capability to reflect the complex dynamics within the path. 

There are various embedding techniques that can be used to augment paths, each suited to specific applications based on the aspects of the data they capture. Different embedding choices highlight different features of the data, making them appropriate for particular tasks \citep{FERMANIAN2021}. In this paper, we focus mainly on three types of embeddings: basepoint augmentation, time augmentation, and the lead-lag transformation.

The basepoint augmentation \citep{kidger2020} involves inserting a fixed initial point to the path data. By anchoring the sequence, basepoint augmentation allows the signature to overcome the translation invariance of the signature.

Time augmentation counteracts the signature's invariance to time reparametrization, by introducing a monotone coordinate that serves as a virtual timestamp to each point in the path. This embeds precise temporal information into the path, making the signature sensitive to the sequence's timing. Besides, this also guarantees the uniqueness of the signatures computed \citep{hambly2010uniqueness}.

To further aid the sequential learning capabilities of the signature, the lead-lag transformation extends the path's dimensionality by interleaving the original path with its lagged version, which enriches the data representation, allowing the signature to capture not only the path's immediate direction but also its past states. \citet{FERMANIAN2021} noted the lead-lag transformation improves a model's predictive capabilities and is beneficial for tasks involving sequential data.

In the proceeding sections, we will use these embedding transformation to prepare our data paths before computing its signature so as to enhance the signatures’ ability to capture and reflect the complex dynamics of the sequences.

\subsection{Signature regression}
One of the many practical applications of signatures are its usage as feature maps in a regression framework. This rich representation allows for the modelling of complex relationships within functional data. The ability of signatures to compactly encode information about paths makes them particularly powerful for handling high-dimensional and complex datasets, enabling the application of regression techniques in scenarios where conventional methods might be inadequate due to the complexity and dimensionality of the data. These are motivated and guaranteed by many results as pointed out by the shuffle identity, the signature approximation theorem \citep{levin2016} and the universal nonlinearity of paths \citep{bonnier2019deep}. 

\begin{definition}[Universal nonlinearity \citep{bonnier2019deep}]
    \label{uninon}
        \textit{Let $f$ be a real-valued continuous function on continuous piecewise smooth paths in $\mathbb{R}^d$ and let $\kappa$ be a compact set of such paths. Assume that $X_0 = 0, \forall X \in \kappa$, let $\epsilon > 0$, there exists a linear functional $L$ such that,}
    \begin{equation}
        |f(X) - L(S(X))| < \epsilon . \nonumber
    \end{equation}

    \end{definition}

\citet{Kiraly2019} then gathered that a linear combination of signature features are capable of approximating continuous functions arbitrarily well, meaning signatures are able to linearise functions of a path, further emphasizing the universality \citep{levin2016} of signatures. To further clarify, consider a classical regression setting with a path $X$  that maps some space into $\mathbb{R}^d$ and $Y$ a response, we have

\[Y=f(X)+\epsilon, \]

\noindent where $f$ is some functional that characterises the relationship between the response and the path. Assuming $f$ is nonlinear, it is then generally challenging to determine the function $f$ accurately due to the increased complexity in parameter estimation, model interpretation, and computational resources required for non-linear regression or possibly neural networks, compared to linear regression. However, if the signature of path $X$ is considered as the predictors instead, we can redefine the relationship with a linear functional $g$ such that

\[Y=g(S(X))+\epsilon.\]

Although in theory linear regression is sufficient for estimating $g$, in practice, regularization techniques are used to handle the multicollinearity of signature features \citep{levin2016} as linear regression models can become unstable when variables are not independent. This instability often results in large variances in the estimated coefficients, making the model unreliable. 

\subsection{The Hyndman-Ullah model with truncated signatures}
We are now prepared to integrate signatures into the Hyndman-Ullah (HU) model, which we will refer to as the HU with truncated signatures (HUts) model. The integration of signatures with functional data models have been documented by \citet{FERMANIAN2022} and \citet{frévent2023functional}. However, these studies primarily applied signature regression in a functional regression framework with scalar responses and functional covariates, where signatures of the functional covariates were used directly. In contrast, the HU model operates more like a functional regression model with functional responses, making the direct application of signature regression more complex. We plan to consider using the signature of cross-sectional data from these functional covariates to apply signature regression at each point throughout the functional domain and then reconstruct a functional response. The HUts model is presented as follows:

\begin{equation}
    \label{eq:HU1}
    y_t(x)=f_t(x)+\sigma_t(x)\epsilon_t, \nonumber
\end{equation}
\begin{equation}
    f_t(x)=\mu(x)+\sum_{k=1}^K \beta_{t,k} Z_k(x) +e_t(x), \nonumber
\end{equation}

\noindent where $Z_k(x)$ is the $k^{th}$ principal component of the signature of the transformed path of age specific rates, and its respective scores $\beta_{t,k}$. 

The smoothing approach of the mortality rates are remained as the original HU model, which is with the constrained weighted penalized regression splines to obtain $\hat{f}_t(x)$. This ensures the model retains the capacity to account for handling heteroskedasticity in mortality data as variability is different across different ages or time periods. Although more advanced smoothing techniques such as bivariate smoothing \citep{Dokumentov2018} have been put forth, they also introduce additional complexity and computational demands.

Additionally, the estimation of the mean function is also kept the same as in the HU model, by averaging all the smoothed curves: 
\begin{equation}
    \hat{\mu}(x)=\frac{1}{n}\sum_{t=1}^{n}\hat{f}_t(x). \nonumber
\end{equation}
This method provides a straightforward and effective way to capture the central tendency of the functional responses. Now we will decompose the centered curves, $f^*_t(x)=\hat{f}_t(x)-\hat{\mu}(x)$ with signature regression over a fine grid of $q$ equally spaced values $x^*_{1}, \dots, x^*_{q} \in[x_1,x_p]$. 

Consider the time series of the centered smoothed mortality rates for age \( x_i \), \( \{f^*_t(x_i)\}_{t=1}^{n} \). Combining the basepoint augmentation, time augmentation, and the lead-lag transformation, the augmented path can be transformed as follows:

\begin{align}
    \tilde{X_i} = \Bigl(&(t_0, 0, 0), (t_1, f^*_1(x_i), f^*_1(x_i)), (t_2, f^*_2(x_i), f^*_1(x_i)), (t_3, f^*_2(x_i), f^*_2(x_i)), \nonumber \\
    & (t_4, f^*_3(x_i), f^*_2(x_i)), \ldots, (t_n, f^*_n(x_i), f^*_n(x_i)) \Bigr) \nonumber,
\end{align}

\noindent where $t$ is an ordered sequence within the interval $[0,1]$ and $t_0 = 0 < t_1 < t_2 < \ldots < t_n = 1$, marking the times of observations. This structured augmentation ensures that each data point $x_i$ is not only represented by its current value but also linked to its immediate past, enhancing the path's dimensionality and capturing both the direction and history within the data, thereby enriching the data representation for signature computation. The transformed time series $\tilde{X}_i$ is then used to compute the truncated signature $S^m(\tilde{X}_i)$.

To estimate $\beta_{t,k}$, we then set up our signature regression framework: \[f^*_t(x_{i})=g_t(S^m(\tilde{X}_i))+\hat{e}_t(x_i)\]
for some linear function $g_t$, which we will use principal component regression to solve. Applying PCA helps by reducing the dimensionality of the signature coefficient feature space to a smaller set of $k$ uncorrelated principal components denoted as $Z_k(x_i)$. This reduction not only simplifies the model but also enhances its stability and efficiency by focusing on the most informative aspects of the dependency within the signatures. Additionally, using PCA aids in orthogonalizing the terms, which simplifies the calculation of forecast variance later and resembles the approaches used in both the original HU and LC models.

For computational simplicity, we define the matrices

\begin{align}
S = \begin{bmatrix}
    S^m(\tilde{X}_1) \\
    \vdots \\
    S^m(\tilde{X}_q)
\end{bmatrix}_{q \times s_d(m)} \;, \hspace{1cm}
Y_t = \begin{bmatrix}
    f^*_t(x_1^{*}) \\
    \vdots \\
    f^*_t(x_q^{*})
\end{bmatrix}_{q \times 1}. \label{eq:Yt_matrix}
\end{align}

We then normalise $S$ to obtain $S^*$ before applying PCA to obtain the orthogonal principal component scores by singular value decomposition, $S^* = U \Sigma V^T$ where the first $K$ principal component scores are the first $K$ columns of $Z = U \Sigma$. Then, to obtain $\hat{\beta}_t$ is equivalent to solving  
\begin{equation}
    \hat{\beta}_t = (Z_K^\top Z_K )^{-1}Z_K^\top Y_t, \nonumber
\end{equation}
where $Z_K$ is a $q \times K$ matrix where the $i^{th}$ row contains the principal component of $S^*$ of order $m$ of age $x_i$, 

By combining everything together we have 
\begin{equation}
    \hat{f}_t(x)=\hat{\mu}(x)+\sum_{k=1}^K \hat{\beta}_{t,k} \hat{Z}_k(x) +\hat{e}_t(x), \nonumber
\end{equation}
where $\hat{Z}_k(x)$ is obtained via interpolation.

\begin{remark}
As time augmentation leads to identical values for the signature coefficients $S^{(1)}, S^{(1,1)}, S^{(1,1,1)},\ldots$, along each row of $S$ in Equation (\ref{eq:Yt_matrix}), certain columns of $S$ become constant. During normalisation, these zero-variance columns are excluded from being centered and scaled, ensuring a well-defined normalized matrix $S^*$. Consequently, when performing PCA on $S^*$, one of the principal components is expected to be constant.
\end{remark}
\subsection{Forecasting with the HUts model}
\subsubsection{Point forecast}
Similar to that of the HU model, the time dependent coefficients are extrapolated by fitting them into univariate time series $\{\hat{\beta}_{t,k}\}_{t=1}^n$. We preserve the choice of using univariate ARIMA models as in the HU model to forecast each coefficient $h$-steps-ahead, $\hat{\beta}_{n+h|n,k}$, for each $k=1,\dots,K$. Subsequently, the forecasted log mortality rates can be calculated with:

\begin{equation}
    \label{Eq: FC}
    \hat{y}_{n+h|n}(x)=\hat{\mu}(x)+\sum_{k=1}^K \hat{\beta}_{n+h|n,k} \hat{Z}_k(x). \nonumber
\end{equation}
Since each term is constructed to be orthogonal to one another, the forecast variance can also be approximated, assuming all sources of error are normally distributed, by summing of the component variances similar to that of the forecast variance of the HU model \citep{HYNDMAN20074942}:
\begin{equation}
    \label{Eq: varHUts}
    \text{Var}(y_{n+h|n}(x)) \approx \hat{\sigma}^2_{\mu}(x) + \sum_{k=1}^K \hat{Z}_k^2(x) u_{n+h|n,k} + v(x) + \sigma^2_{n+h|n}(x),
\end{equation}

\noindent where $\hat{\sigma}^2_{\mu}(x)$ is the variance obtained from estimating the mean function, $ u_{n+h|n,k}$ is the variance from the time series used to forecast $\hat{\beta}_{n+h|n,k}$, $ v(x)$ are model fit errors estimated by averaging $\hat{e}_t(x)$, and finally $\sigma^2_{n+h|n}(x)$, the observational variance.

\subsubsection{Interval forecast}
Obtaining prediction intervals is essential because they provide a measurable indication of the uncertainty in forecasts. Prediction intervals quantify the range within which future observations are likely to occur, at a specified confidence level. This allows evaluation of the reliability of predictions when model uncertainty is significant, as different models may offer varying forecasts performance. \citet{chatfield} emphasized the importance of prediction intervals for evaluating future uncertainty, preparing for various scenarios, comparing different forecasting methods, and exploring outcomes under different assumptions. By providing a range of possible outcomes, interval forecasts allow policymakers and researchers to prepare for a variety of future scenarios, thus enhancing the reliability of their strategies against unexpected changes. For example, understanding the variability in future mortality rates can profoundly impact the structuring of insurance \citep{HYNDMAN2009}. Interval forecasts helps insurers assess the risk associated with different demographic scenarios, which is then translated into premiums to reflect the uncertainty of their projections. This further frames the concept of interval forecasts as a fundamental component in the field of demography.

Constructing prediction intervals is straightforward under the assumption that all sources of error are uncorrelated and normally distributed. Using the standard normal distribution, these intervals can be easily derived by applying quantile values to scale the forecast standard deviation obtained in Equation (\ref{Eq: varHUts}) which is straightforward to implement. However, its accuracy heavily depends on the validity of the normality assumption. If the normality assumption is violated, the prediction intervals may not be accurate, suggesting that a nonparametric bootstrap procedure would be more practical.

\subsubsection{Bootstrap prediction interval}
Since the HUts model, like the wHU model, do not satisfy the normality assumption (see \ref{AppendixA} in the Appendix), we employ the same method used to construct bootstrap prediction intervals for the wHU model in \citet{HYNDMAN2009} to construct prediction intervals for the HUts model.

The procedure begins by utilizing univariate time series models to generate multi-step-ahead forecasts for the scores of principal components, $\{\beta_{t,k}\}_{t=1}^n$. Subsequently, for $t = h+1, \dots, n$, we calculate the forecast errors for $h$-step-ahead forecasts as:
\begin{equation}
    \hat{\xi}_{t,h,k} = \hat{\beta}_{t,k} - \hat{\beta}_{t|t-h,k}. \nonumber
\end{equation}

To create a bootstrap sample of $\beta_{n+h,k}$, we randomly resample these forecast errors with replacement, designated as $\hat{\xi}_{*,h,k}^{(l)}$, and add them to the $h$-step-ahead forecast, $\hat{\beta}_{n+h|n,k}$, yielding:
\begin{equation}
    \hat{\beta}_{n+h|n,k}^{(l)} = \hat{\beta}_{n+h|n,k} + \hat{\xi}_{*,h,k}^{(l)}, \nonumber
\end{equation}

\noindent for $l=1, \dots, L$.

The model's residual, $\hat{e}_{n+h|n}^{(l)}(x)$ is bootstrapped from the residuals $\{\hat{e}_1(x), \ldots, \hat{e}_n(x)\}$ by sampling with replacement. Similarly, smoothing errors $\hat{\epsilon}_{n+h,i}^{(l)}$ can be bootstrapped by resampling with replacement from $\{\hat{\epsilon}_{1,i}, \ldots, \hat{\epsilon}_{n,i}\}$.

When constructing the $L$ variants for $\hat{y}_{n+h|n}^{(l)}(x)$, all possible components of variability are combined, assuming they are uncorrelated to each other by using:
\begin{equation}
\hat{y}_{n+h|n}^{(l)}(x_i) = \hat{\mu}(x) + \sum_{k=1}^K \hat{\beta}_{n+h|n,k}^{(l)} \hat{Z}_k(x_i) + \hat{e}_{n+h|n}^{(l)}(x_i) + \hat{\sigma}_{n+h}(x_i)\hat{\epsilon}_{n+h,i}^{(l)}. \nonumber
\end{equation}
Subsequently, the $(1-\alpha)100\%$ prediction intervals, $[l_h(x),u_h(x)]$ can then be obtained from the $(1-\alpha/2)$ and $(\alpha/2)$ quantiles of these bootstrap variants. We also consider the adjustment made to the bootstrap prediction intervals in \citet{HYNDMAN2009}:
\begin{equation}
    [0.5\{l_h(x)+u_h(x)\}-\{u_h(x)-l_h(x)\}p(x),0.5\{l_h(x)+u_h(x)\}+\{u_h(x)-l_h(x)\}p(x)] \nonumber
\end{equation}
to account for the uncertainty in our forecast by calibrating the intervals using one-step-ahead forecast errors. Here, $p(x)=d(x)/[u_1(x)-l_1(x)]$, where $d(x)$ is the empirical differences between the $(1-\alpha/2)$ and $(\alpha/2)$ quantiles of the in-sample one-step-ahead forecast errors $\{\hat{f}_{t+1}(x)-\hat{f}_{t+1|t}(x); t=r,\cdots,n-1\}$, and $r$ is the smallest number of observations used to fit the model.

\section{Empirical analysis}

\FloatBarrier
\subsection{Data description}
This study utilised datasets from 12 different countries, all sourced from the \citet{HMD}. Of the 12 countries considered, data spanning all available years up to 2015 were extracted, with age groups ranging from zero to a carefully determined maximum age to avoid zero or missing values in elderly data. For ages beyond this established maximum, they were combined into an upper age group. In addressing intermediate missing values within the data for Denmark, Finland and Norway, a uniform approach was applied. A constant, represented by the smallest death rate value in the dataset, was added to each death rate prior to taking its logarithm. This adjustment facilitated a consistent treatment of missing values across the dataset. The selected countries are tabulated in Table \ref{tab:data_descn}, along with the commencing years and age groups.

\begin{table}[htbp]
\centering
\caption{Mortality data by country}
\label{tab:data_descn}
\begin{tabular}{@{}lll@{}}
\toprule
Country       & Commencing Year & Age Range \\ \midrule
Australia     & 1921            & 0-100+    \\
Belgium       & 1920            & 0-100+    \\
Bulgaria      & 1947            & 0-100+    \\
Denmark       & 1899            & 0-99+     \\
Finland       & 1899            & 0-96+     \\
France        & 1899            & 0-100+    \\
Ireland       & 1950            & 0-100+    \\
Italy         & 1899            & 0-100+    \\
Japan         & 1947            & 0-100+    \\
Netherlands   & 1899            & 0-99+    \\
Norway        & 1899            & 0-100+   \\
United States & 1933            & 0-100+    \\ \bottomrule
\end{tabular}
\end{table}

\FloatBarrier
\subsection{Point forecast}
The log mortality rates of 12 countries were forecasted with the HUts model using a truncation order $m=2$ (see \ref{AppendixB} in the Appendix). It is therefore intuitive to compare the forecast accuracy of the proposed method with the HU model, HUrob model, and the wHU model, which are commonly used on mortality rates of a single population. The selection of the number of principal components, $K$ can impact the forecasting performance of the HUts model and the HU model variants. They can be chosen based off a threshold function that measures the proportion of variance explained by each principal component, or by minimizing the integrated squared forecast error \citep{HYNDMAN20074942}. However, \citet{HYNDMANbooth2008} noted that the HU models are relatively insensitive to the choice of $K$ as long as it is sufficiently large, and that setting $K=6$ is more than adequate to capture the essential features of the data. For consistency across all models, we adopt $K=6$ for both the HU model variants and the proposed HUts model. The models were fitted using an expanding window approach, incorporating data from the respective commencing years of each country as outlined in Table \ref{tab:data_descn}. The initial forecasting period covers the most recent two decades which ends at 2015. We fit this data using both models and produce forecasts up to ten years ahead. The accuracy of these forecasts are then evaluated. Subsequently, the fitting period is then extended by a year and the forecasts for the same horizons are recalculated. This procedure of expanding the training window continues until it includes data up to the year 2014.

In order to gauge the accuracy of the point forecast of the HUts model, the mean squared errors (MSE) are calculated and compared with those of the forecasts of the HU model variants. MSE quantifies the average squared difference between the predicted and observed values, providing a measure of the overall accuracy and precision of predictions with lower values indicating better forecasting performance. The MSE is computed using:
\begin{equation}\label{eq:mse_mape_h}
        MSE(h) = \frac{1}{pq}\sum^q_{t=1}\sum^p_{i=1}(y_t(x_i)-\hat{y}_{t|t-h}(x_i))^{2},
\end{equation}
where $h$ is the forecasting horizon, $p$ is the number of ages, $q$ is the number of years considered in the forecasting period, $y_t(x_i)$ is the observed log mortality, and $\hat{y}_{t|t-h}(x_i)$ is the predicted value.


\begin{figure}[htbp]
    \centering
    
    \begin{subfigure}{\textwidth}
        \centering
        \subcaption{}
        \includegraphics[width=0.75\textwidth]{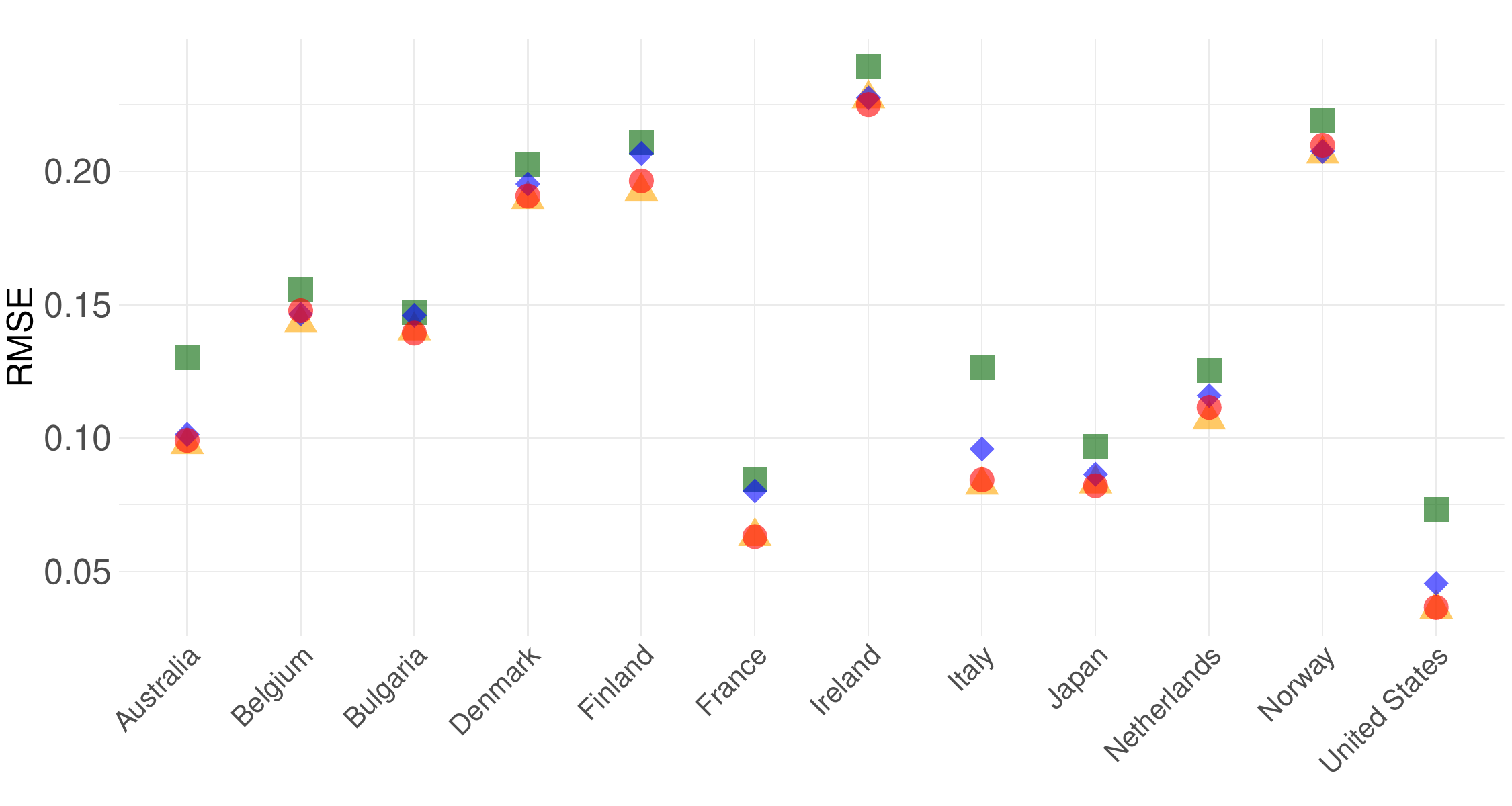}
        \label{fig:h1}
    \end{subfigure}
    
    
    \begin{subfigure}{\textwidth}
        \centering
        \subcaption{}
        \includegraphics[width=0.75\textwidth]{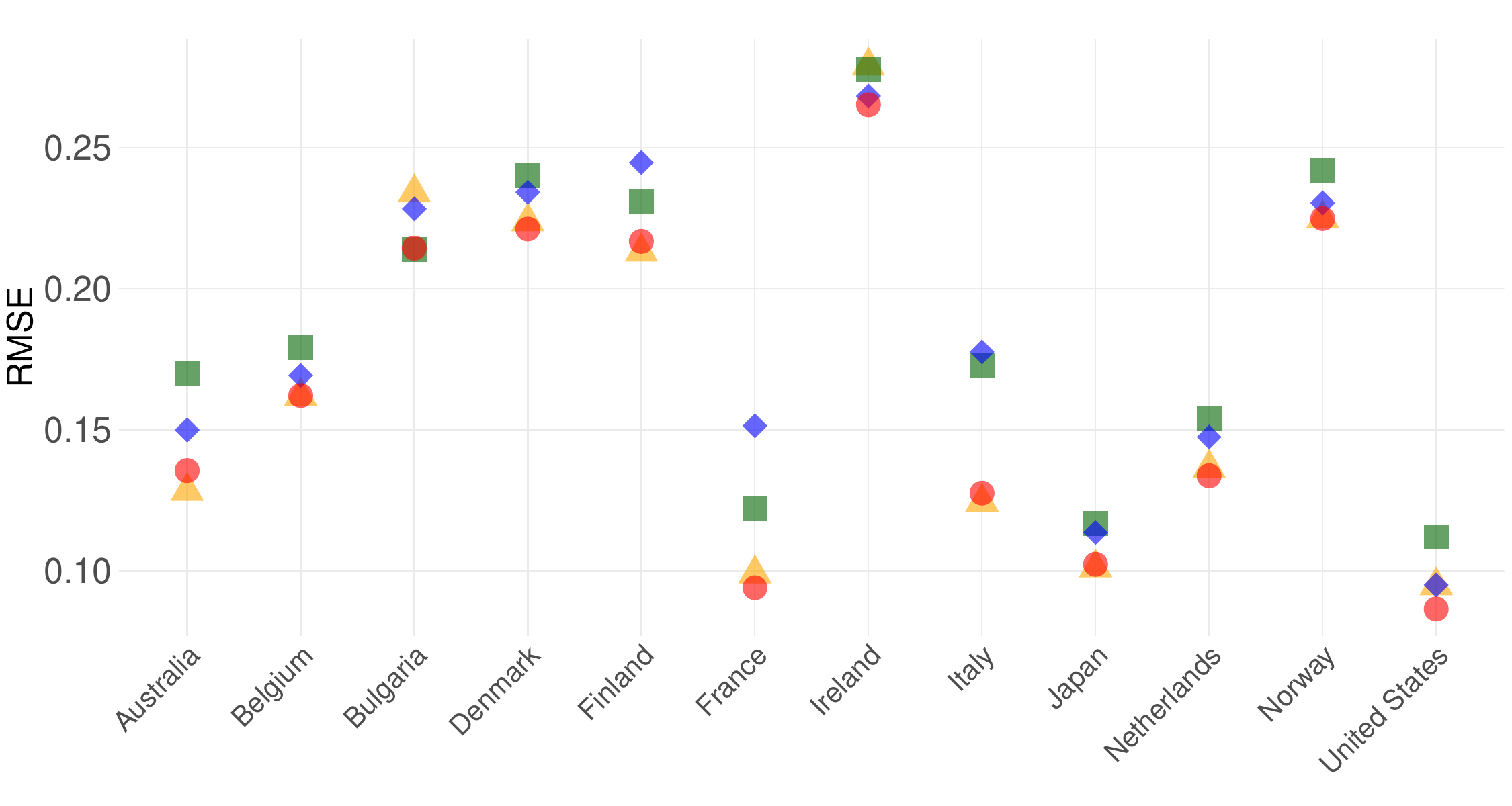}
        \label{fig:h5}
    \end{subfigure}
    

    \begin{subfigure}{\textwidth}
        \centering
        \subcaption{}
        \includegraphics[width=0.75\textwidth]{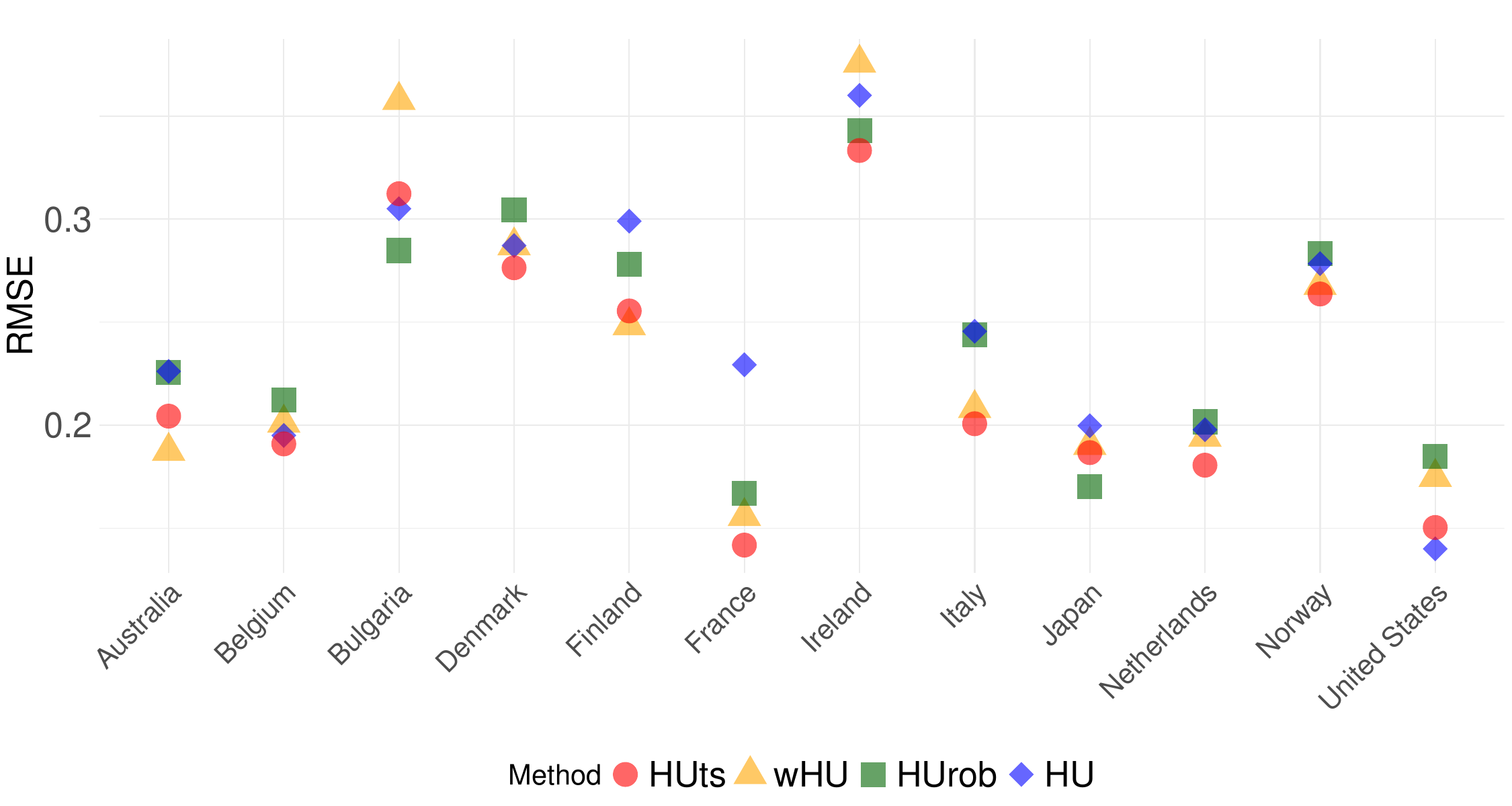}
        \label{fig:h10}
    \end{subfigure}
    
    \caption{Root MSE of (a) one, (b) five, and (c) ten-step-ahead point forecasts of log mortality rates by model and country}
    \label{fig:msescatterplot}
\end{figure}

Figure \ref{fig:msescatterplot} presents the plot of the root MSE of $h=1,5,10$ forecasts across the different models. The square roots of the MSE values are plotted instead of the MSE themselves to ensure the plot scale accommodates all data. Readers are directed to \ref{AppendixD} in the Appendix for the tabulated MSE values of all countries. 
The results indicate that while the wHU model often achieves the lowest MSE values for one-step-ahead forecasts, the HUts model still performs competitively, frequently achieving values close to the best-performing models. In comparison, the HU model shows higher error rates, followed by the HUrob model.

For five-step-ahead forecasts, the HUts model demonstrates more competitive performance. Figure \ref{fig:msescatterplot} (b) shows that overall the HUts model achieves lower error metrics compared to the other models, suggesting an improvement in the HUts model's forecasting ability as the forecast horizon increases.

The results for ten-step-ahead forecasts, shown in Figure \ref{fig:msescatterplot} (c), further confirms the strong performance of the HUts model. It consistently achieves lower or comparable MSE, reinforcing its effectiveness in long-term mortality rate forecasting. In most countries, the HUts model has lower MSE values, with exceptions to a few cases such as in the United States, where the HU model occasionally performs better. This consistent performance across a majority of countries highlights the robustness of the HUts model in capturing long-term trends in mortality rates.

Across different forecasting horizons, the best model is generally a choice between the HUts and wHU models, as these models tend to have the lowest MSE values. The wHU model performs better at shorter forecast horizons, as seen in Figure \ref{fig:msescatterplot} (a). However, as the forecasting horizon increases, the HUts model demonstrates improved performance and eventually surpasses the wHU model. This shift indicates that the HUts model's accuracy and reliability become more pronounced with longer-term forecasts.

Notably, in the cases of France and Ireland, the HUts model performs consistently well across all forecasting horizons, maintaining lower MSE values compared to other models. Conversely, for Australia, the HUts model does not appear to be the most favorable at any point across the forecast horizons, with other models, such as the wHU model, often yielding better results. The following section will take a closer look at the mortality rate forecasts for France and Australia, exploring the nuances and potential reasons behind the differing performances of the HUts model against the HU model variants.

\FloatBarrier
\subsubsection{A closer look at the French and Australian mortality data}
\begin{figure}[htbp]
\centering
\includegraphics[width=0.75\linewidth]{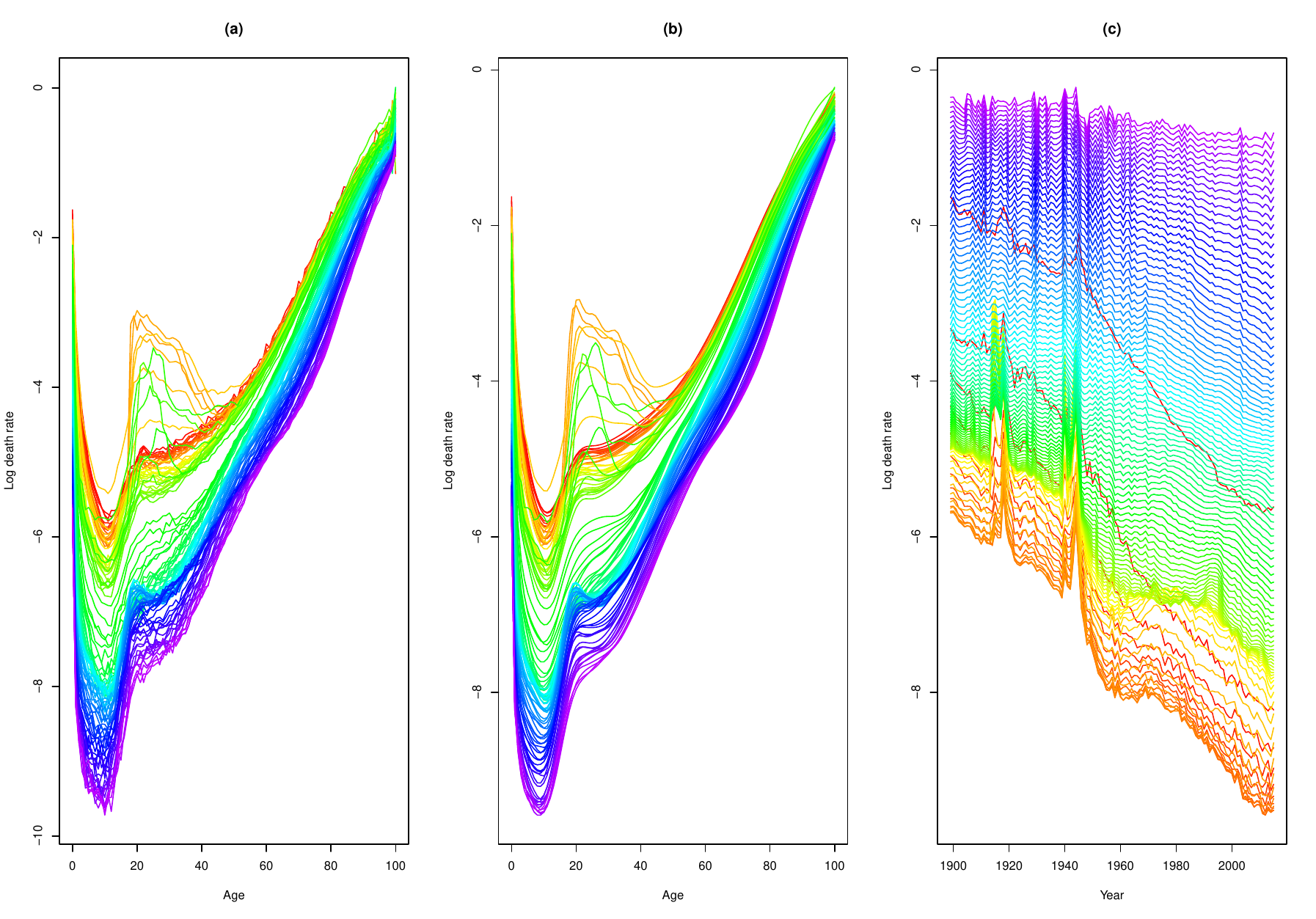}
\caption{France (1899 to 2015); (a) Observed log mortality rates; (b) Smoothed log mortality rates; (c) Times series for ages}
\label{fig:frmort}
\end{figure}

\begin{figure}[htbp]
\centering
\includegraphics[width=0.75\linewidth]{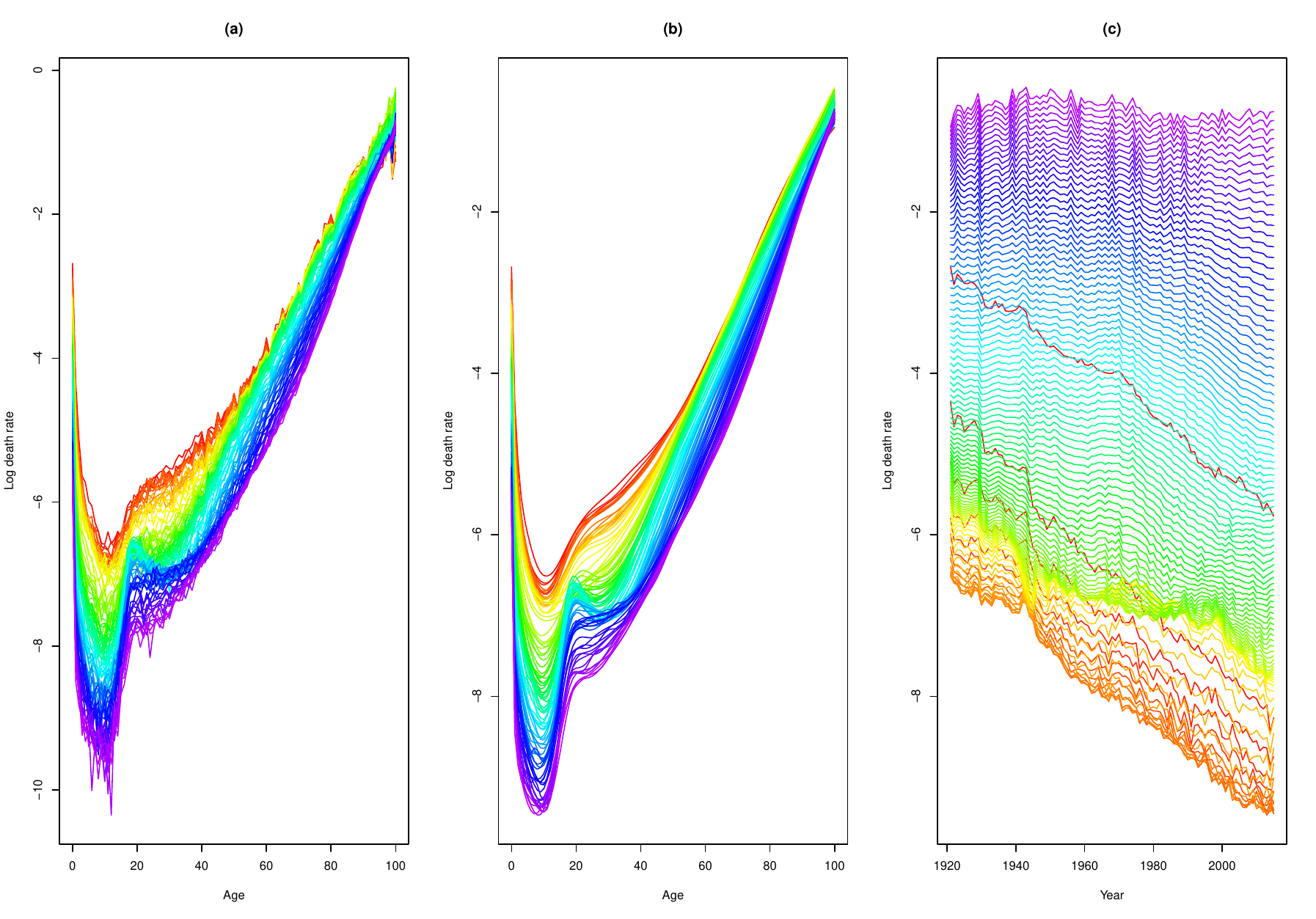}
\caption{Australia (1921 to 2015); (a) Observed log mortality rates; (b) Smoothed log mortality rates; (c) Times series for ages}
\label{fig:ausmort}
\end{figure}

To assess the robustness of the HUts model, we examine the mortality rates of France and Australia more closely. Figures \ref{fig:frmort} (a) and \ref{fig:ausmort} (a) include the observed log mortality rates, and Figures \ref{fig:frmort} (b) and \ref{fig:ausmort} (b) show the smoothed log mortality rates with red representing the earlier years. Figures \ref{fig:frmort} (c) and \ref{fig:ausmort} (c) the smoothed log mortality rates viewed as a time series with red representing the earlier ages. France's mortality data, notably influenced by periods of war and disease outbreaks, presents distinct outliers which are evident in the volatility observed in their mortality trends, as shown in Figure \ref{fig:frmort} which are most visible around the 1920s and 1940s. This provides an oppurtunity to test for the HUts model's ability to handle anomalies in data. In contrast, Australia's mortality data, starting from 1921 appears to have a smoother trend, without any significant outliers as depicted in Figure \ref{fig:ausmort}.

\begin{figure}[htbp]
\centering
\includegraphics[width=0.75\linewidth]{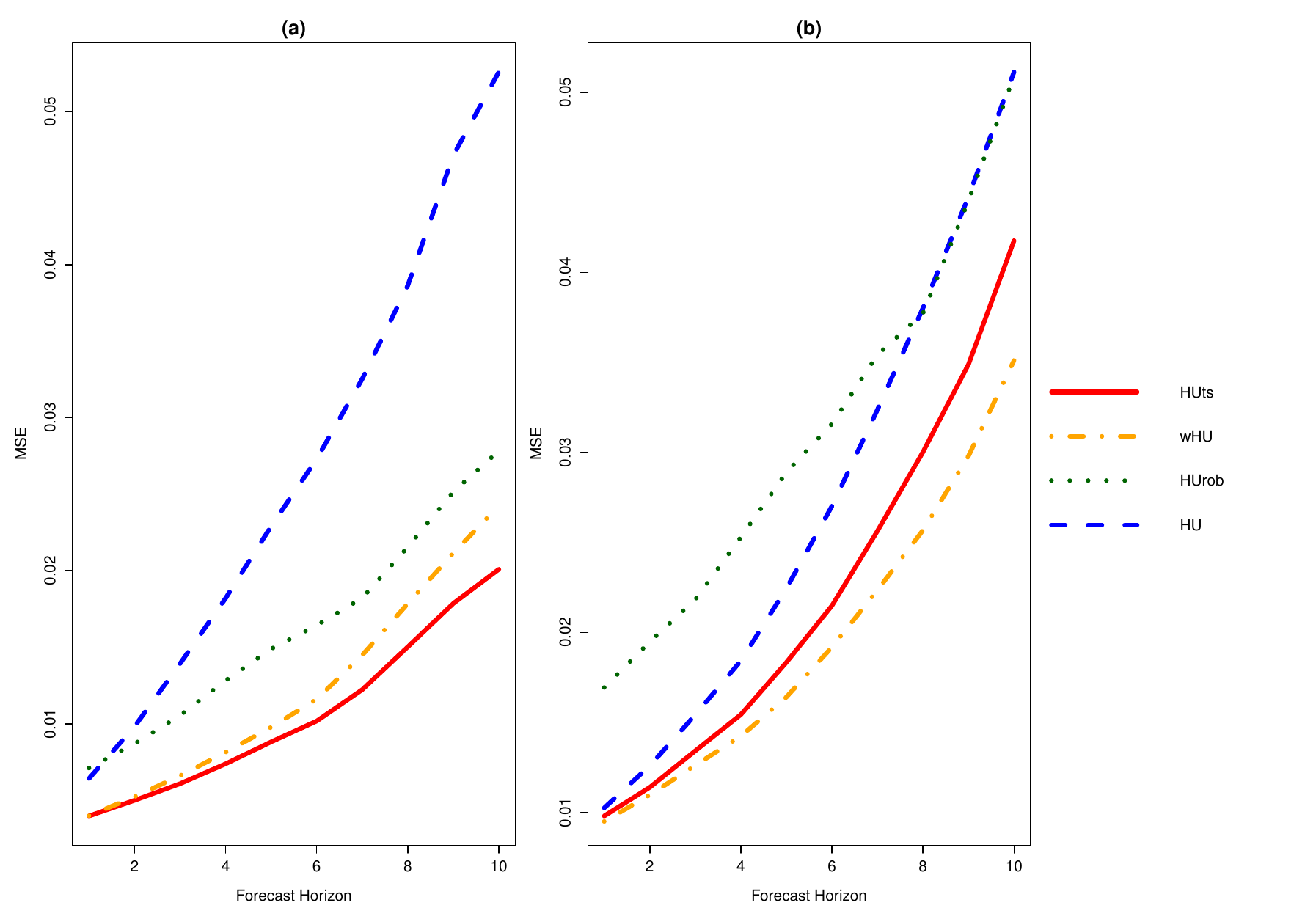}
\caption{MSE plots for (a) French  and (b) Australian mortality of the HUts, wHU, HUrob, and HU models}
\label{fig: frME}
\end{figure}

The ability of the HUts model to effectively handle anomalies in French mortality is illustrated in Figure \ref{fig: frME} (a), where the model demonstrates a substantial reduction in MSE across various forecast horizons when compared to the HU model. This significant improvement shows the HUts model's robustness against data irregularities, making it a reliable choice for forecasting in scenarios with potential outliers. Additionally, the HUts model showcases lower MSE than both the HUrob and wHU models, further affirming its superior performance in handling French mortality. Conversely, Australia presents a more stable dataset with mortality rates following a consistent trend and lacking significant outliers. The MSE plot for Australia in Figure \ref{fig: frME} (b) shows less pronounced differences compared to France. While the HUts model demonstrates superior performance over the HU and HUrob models, affirming its effectiveness even in more uniform datasets, the wHU model performs better than the HUts model. This suggests that the wHU model might be more suited for datasets with consistent trends and fewer irregularities. This comparative analysis highlights the importance of selecting appropriate models based on the characteristics of the dataset, with the HUts model excelling in more variable datasets and the wHU model in more stable ones. The MSE plots of the remaining countries can be found in \ref{AppendixE} in the Appendix.

To gain deeper insights into the model performance, the MSE values are considered across the age spectrum. Instead of using Equation (\ref{eq:mse_mape_h}), the MSE values are averaged only over the years of the forecasting period. This approach will allow us to observe how the models perform at different age groups, providing a more granular understanding of the forecast accuracy. The MSE across ages are computed with the following:

\begin{equation}
    \label{eq:mse_mape_me_age}
    MSE(h,x_i) = \frac{1}{q}\sum_{t=1}^{q}(y_t(x_i)-\hat{y}_{t|t-h}(x_i))^2. \nonumber
\end{equation}

\begin{figure}[htbp]
\centering
\includegraphics[width=0.75\linewidth]{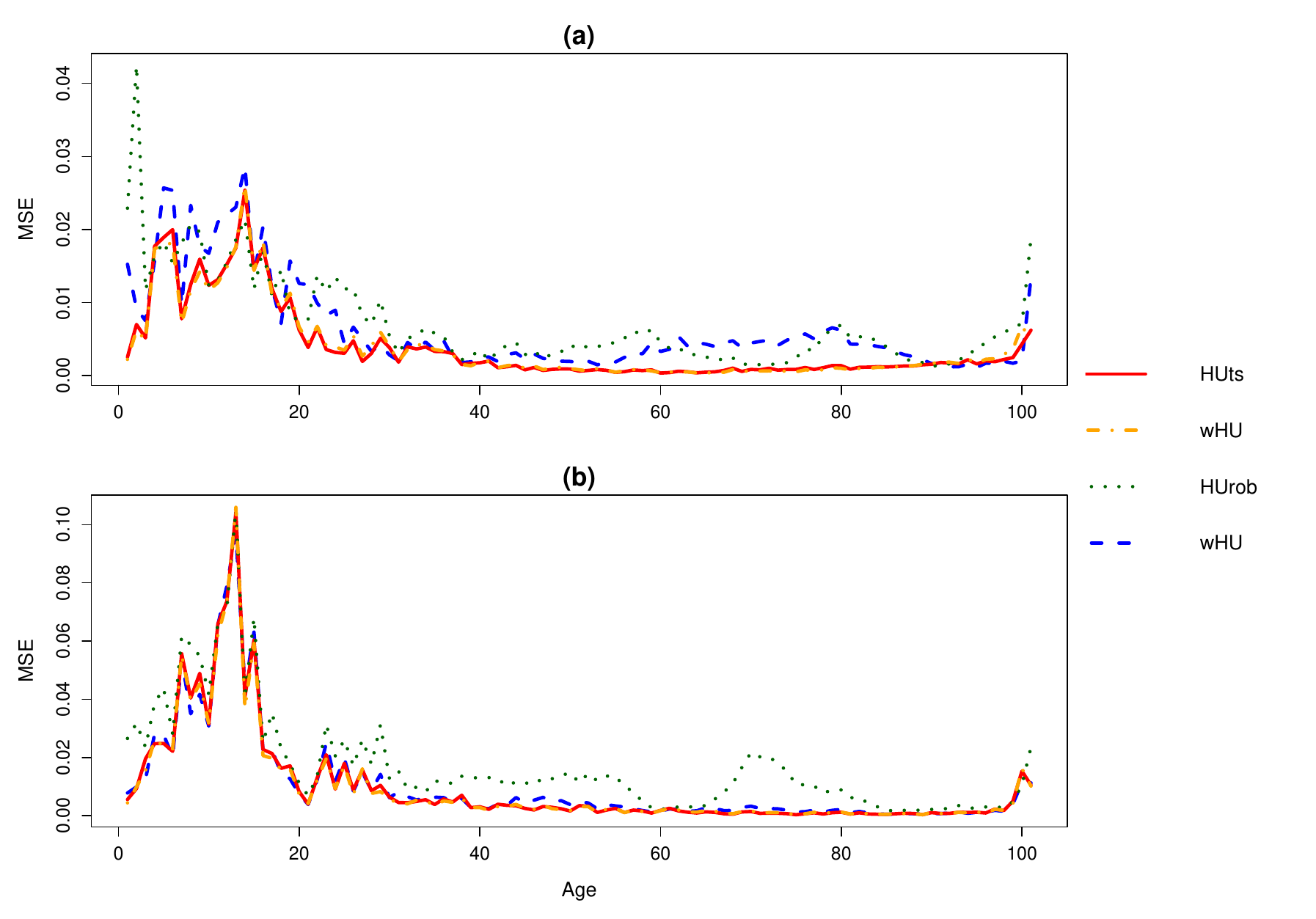}
\caption{One-step-ahead forecast MSE averaged over years in forecasting period of (a) French and (b) Australian mortality for the HUts, wHU, HUrob, and HU models}
\label{fig: frMEh1}
\end{figure}

\begin{figure}[htbp]
\centering
\includegraphics[width=0.75\linewidth]{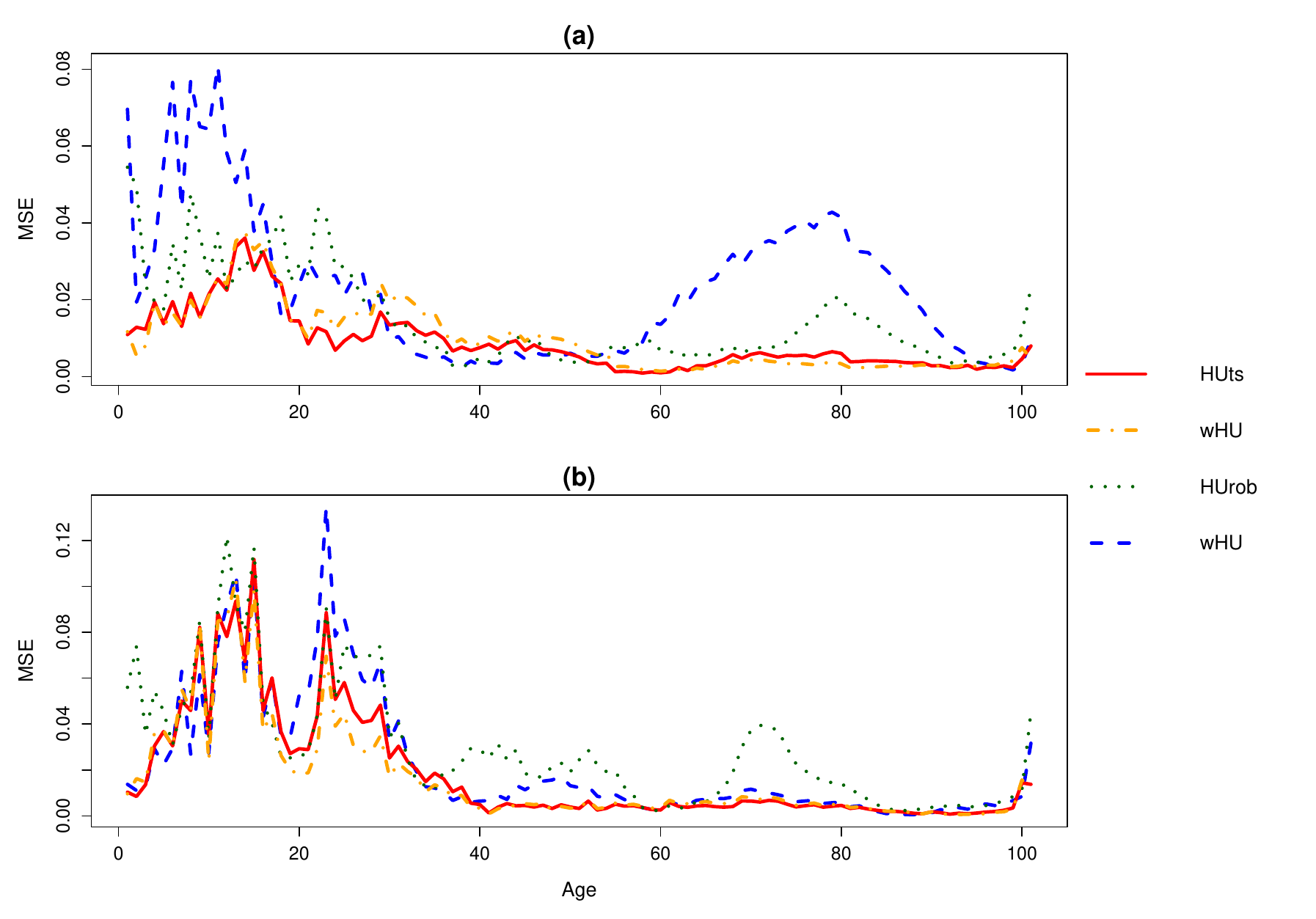}
\caption{Five-step-ahead forecast MSE averaged over years in forecasting period of (a) French and (b) Australian mortality for the HUts, wHU, HUrob, and HU models}
\label{fig: frMEh5}
\end{figure}

\begin{figure}[htbp]
\centering
\includegraphics[width=0.75\linewidth]{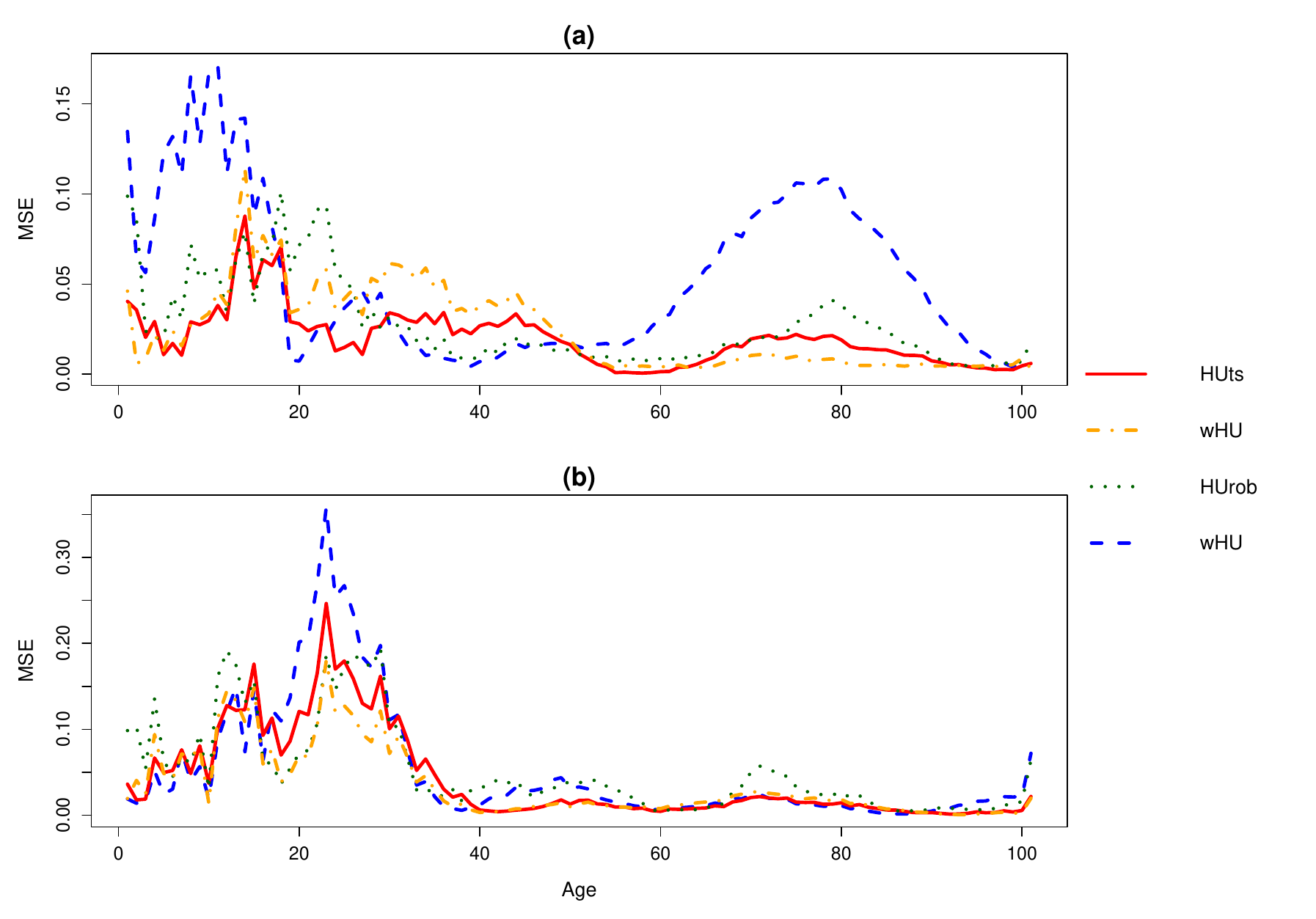}
\caption{Ten-step-ahead forecast MSE averaged over years in forecasting period of (a) French and (b) Australian mortality for the HUts, wHU, HUrob, and HU models}
\label{fig: frMEh10}
\end{figure}

Upon a closer inspection of the MSE across different age groups of the French mortality data in Figures \ref{fig: frMEh1} (a), \ref{fig: frMEh5} (a), and \ref{fig: frMEh10} (a), the HUts model generally surpasses every other model in terms of forecasting performance. For one-step-ahead forecasts (Figure \ref{fig: frMEh1} (a)), the HUts and wHU models show similar performance in terms of MSE, with both models outperforming the HU and HUrob models substantially. For five-step-ahead forecasts (Figure \ref{fig: frMEh5} (a)), the HUts model continues to perform well, maintaining lower MSE values across most age groups compared to the other models. The HUts model shows improved performance at middle ages (30-50) over the wHU model, while the HU and HUrob models, although better at these ages, struggle during early and older ages, contributing to their higher overall error metric. For ten-step-ahead forecasts (Figure \ref{fig: frMEh10} (a)), the HUts model's performance remains robust, achieving lower MSE values across most age groups. However, it still faces competition from the wHU model, which performs better in certain age ranges. Overall, these observations suggest that while the HUts model performs exceptionally well, particularly as the forecast horizon increases, the wHU model also shows strong performance, especially at shorter forecast horizons. The choice between these models often depends on the specific age group and forecasting horizon, with the HUts model becoming more favorable as the horizon increases.

Similarly, MSE for the Australian mortality forecasts in Figures \ref{fig: frMEh1} (b), \ref{fig: frMEh5} (b), and \ref{fig: frMEh10} (b) highlight the competitive nature of the HUts model, particularly at shorter forecast horizons. For one-step-ahead forecasts (Figure \ref{fig: frMEh1} (b)), the HUts model performs comparably to the wHU model, with both showing nearly identical MSE values. This indicates that the HUts model is effective at short-term forecasting. At the five-step-ahead horizon (Figure \ref{fig: frMEh5} (b)), the HUts model continues to perform well but begins to show higher MSE values in the 20-40 age range compared to the wHU model. This underperformance in mid-age groups becomes more evident, reflecting some potential biases that start to emerge at this forecast horizon. By the ten-step-ahead forecasts (Figure \ref{fig: frMEh10} (b)), the HUts model, although still competitive, shows further increased MSE values around ages 20-40. This trend highlights the model's underestimation (see \ref{AppendixF} of the Appendix) in this specific age group, significantly affecting the overall error metric. Despite this, the HUts model maintains reasonable performance across other age groups, yet the wHU model generally achieves better results at these longer horizons, suggesting that the wHU model might be better suited for datasets with more consistent trends and fewer irregularities.

The findings of French and Australian mortality data present a comparative analysis of the HUts model's performance under differing demographic conditions and historical events influencing mortality data. While the HUts model is able to perform well in stable demographic conditions, its advantages are particularly noticeable when confronting data with potential outliers, making it a robust tool for forecasting when compared with the HU model variants.

\FloatBarrier
\subsection{Interval forecast} 

To evaluate the interval forecasts, the empirical coverage probability is calculated using the following expression (Hyndman and Shang, 2009):
\begin{equation}
    \frac{1}{qph} \sum_{t=1}^q \sum_{j=1}^h \sum_{i=1}^p \mathbf{1}\left(\hat{y}_{t+j|t}^{(\alpha/2)}(x_i) < y_{t+j}(x_i) < \hat{y}_{t+j|t}^{(1-\alpha/2)}(x_i)\right), \nonumber
\end{equation}

\noindent where $\mathbf{1}(\cdot)$ is the indicator function, and $\hat{y}_{t+j|t}^{(\alpha/2)}(x_i)$ is the $\alpha/2$-quantile from the bootstrapped samples. It quantifies whether the intervals are adequately wide to encompass the actual observations at a specified nominal coverage of $(1-\alpha)100\%$. 95\% prediction intervals of the HUts model and the wHU model are constructed for all the 12 countries and tabulated in Table \ref{table:95boot}. The minimum observations required to obtain the one-step-ahead forecast errors used for the adjustment, $r$ is set to 30.

\begin{table}[htbp]
\tabcolsep=0pt%
\caption{Empirical coverage probabilities of bootstrapped prediction intervals}\label{table:95boot}
\begin{tabular*}{\textwidth}{@{\extracolsep{\fill}}lcccc@{}}
\toprule
 & \multicolumn{2}{c}{HUts} & \multicolumn{2}{c}{wHU} \\
\cmidrule(lr){2-3} \cmidrule(lr){4-5}
       & 95\% & Adjusted 95\% & 95\% & Adjusted 95\% \\
\midrule
Australia      & 92.70\% & 94.77\% & 88.30\% & 89.22\% \\
Belgium        & 97.36\% & 96.09\% & 91.05\% & 81.13\% \\
Bulgaria       & 91.89\% & 85.44\% & 82.51\% & 79.47\% \\
Denmark        & 92.99\% & 95.84\% & 78.06\% & 91.17\% \\
Finland        & 97.39\% & 98.80\% & 94.68\% & 98.97\% \\
France         & 100.00\% & 100.00\% & 99.95\% & 100.00\% \\
Ireland        & 92.81\% & 82.62\% & 73.54\% & 80.54\% \\
Italy          & 99.84\% & 99.94\% & 99.42\% & 99.86\% \\
Japan          & 97.08\% & 97.39\% & 95.84\% & 79.40\% \\
Netherlands    & 99.61\% & 99.95\% & 98.15\% & 99.94\% \\
Norway         & 94.30\% & 97.98\% & 88.26\% & 98.15\% \\
United States  & 96.60\% & 99.11\% & 92.39\% & 85.70\% \\
\bottomrule
\end{tabular*}
\end{table}

\begin{figure}[htbp]
\centering
\includegraphics[width=0.75\linewidth]{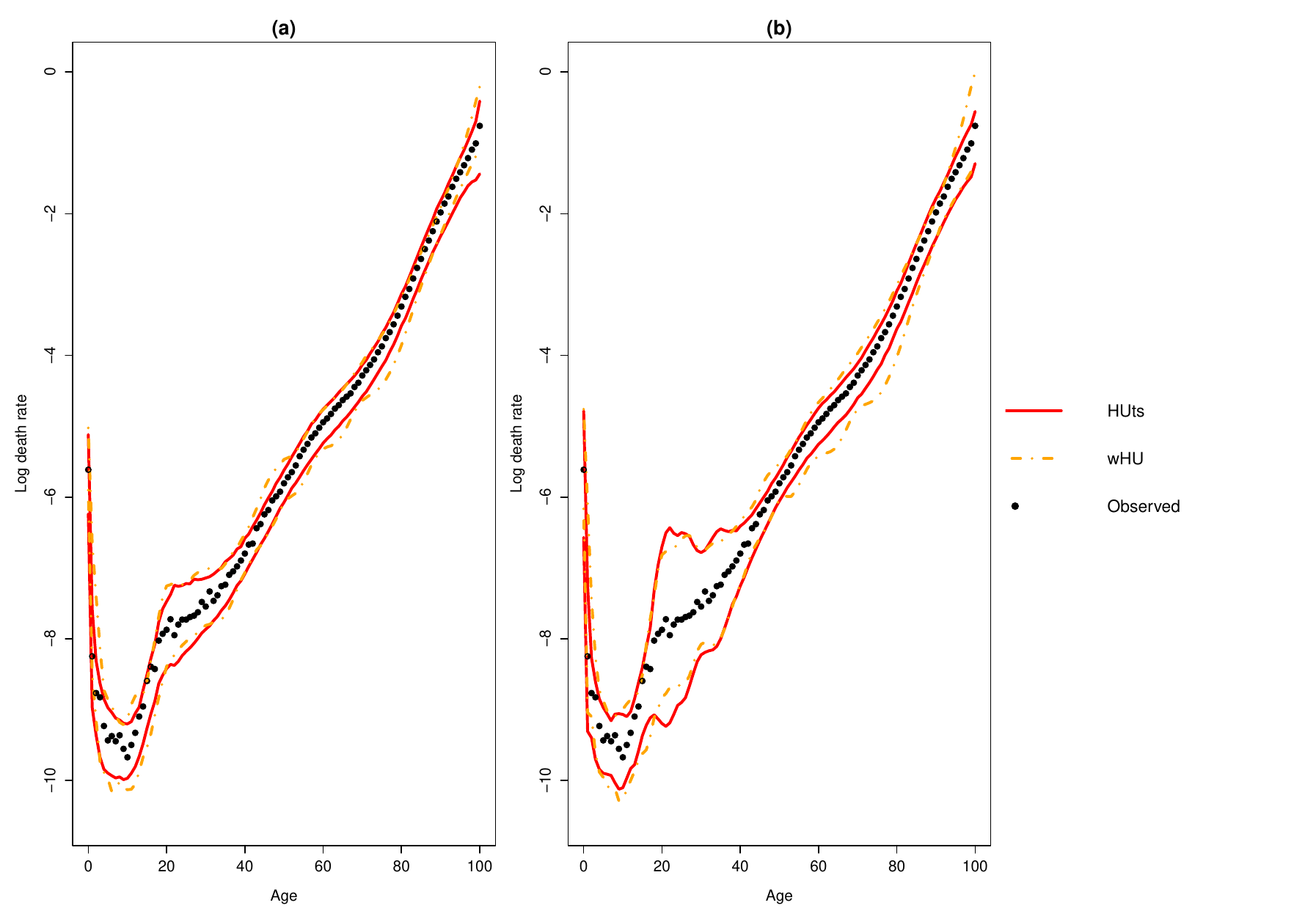}
\caption{Bootstrapped (a) 95\% (b) adjusted 95\% prediction intervals of one-step-ahead forecast for French mortality (1899-2014) for the HUts model and the wHU model}
\label{fig: frPI}
\end{figure}

\begin{figure}[htbp]
\centering
\includegraphics[width=0.75\linewidth]{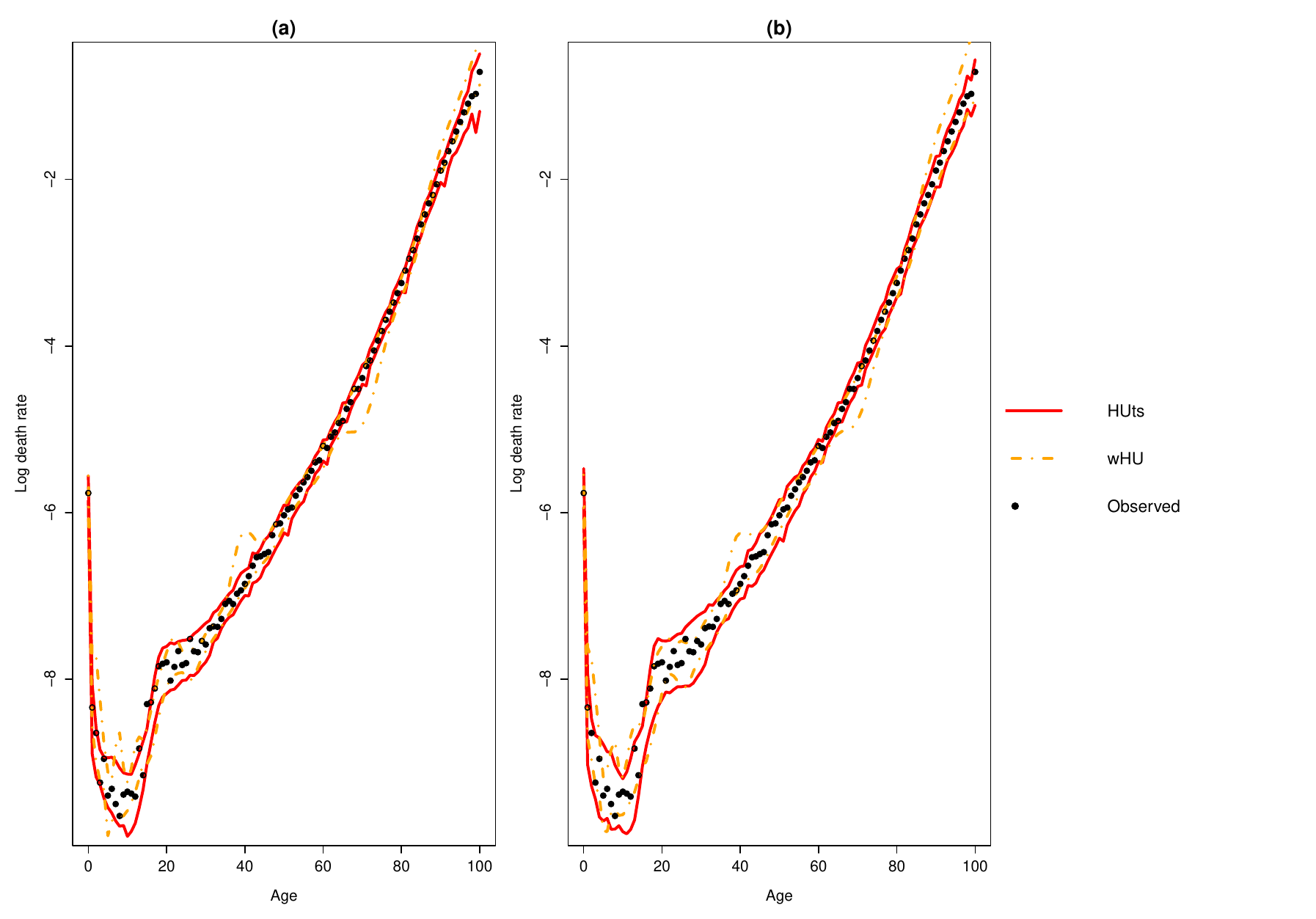}
\caption{Bootstrapped (a) 95\% (b) adjusted 95\% prediction intervals of one-step-ahead forecast for Australian mortality (1921-2014) for the HUts model and the wHU model}
\label{fig: ausPI}
\end{figure}

The bootstrapping procedure generally performs well in obtaining prediction intervals that are close to the nominal 95\% coverage level. Across various countries and forecasting methods, the unadjusted 95\% prediction intervals yield empirical coverage probabilities that align reasonably well with the expected nominal level. These results demonstrate that the bootstrap approach reliably captures the underlying variability in mortality data, ensuring that the prediction intervals accurately achieve the desired 95\% coverage.

However, when applying the \citet{HYNDMAN2009} adjustment to these prediction intervals, the effectiveness of the adjustment varies across different countries and methods. This adjustment aims to refine the intervals based on one-step-ahead forecast errors, potentially bringing the empirical coverage probabilities closer to the nominal level. Yet, its impact is not uniformly beneficial and depends on the specific characteristics of the mortality data and the forecasting models used.

In the case of Australia, both the prediction intervals of the HUts and wHU models show improvements in coverage probabilities after the adjustment. For the HUts method, the empirical coverage increased from 92.70\% to 94.77\%, moving closer to the nominal 95\% level. Similarly, for the wHU method, the coverage improves from 88.30\% to 89.22\%. This indicates that the adjustment effectively enhances the reliability of the prediction intervals in these instances. A similar positive effect is observed for Denmark with the wHU model, where the coverage probability significantly increases from 78.06\% to 91.17\% after adjustment, as well as from 92.99\% to 95.84\% with the HUts model.

On the other hand, the adjustment does not always lead to better coverage probabilities. For example, in Belgium using the wHU method, the coverage probability decreases from 91.05\% to 81.13\% after adjustment, moving further away from the nominal level. Similarly, in Bulgaria with the HUts method, the coverage dropped from 91.89\% to 85.44\% post-adjustment. These instances indicate that the adjustment may not universally improve the prediction intervals possibly due to the nature of the forecast errors of the fitted models.

A notable case is France, where the empirical coverage probabilities reaches 100\% for both unadjusted and adjusted intervals with the HUts model. This phenomenon is caused by the prediction intervals absorbing a significant amount of uncertainty from the forecasts, resulting in wider intervals as seen in Figure \ref{fig: frPI}. For the unadjusted prediction intervals, the uncertainty arises from bootstrapping the model errors. Due to the presence of outliers in the French mortality data, these model errors are larger. When the prediction intervals are adjusted, they incorporate additional variability from the in-sample one-step-ahead forecast errors. These forecast errors are also inflated by outliers, resulting in even wider intervals. This further widening makes the predictions more conservative to ensure coverage despite the high variability in the data. The same effect is observed with the wHU model, where the prediction intervals are similarly wide which suggests that outliers can impact the bootstrapping approach, potentially reducing the reliability of the intervals.

In contrast, when examining the prediction intervals for Australia in Figure \ref{fig: ausPI}, they are significantly narrower than those for France. Australia’s mortality data is more stable, with fewer outliers, leading to less uncertainty in the forecasts. Consequently, the prediction intervals are narrower. This comparison highlights the point that uncertainty within mortality data leads to wider bootstrapped prediction intervals, as seen in France, while more stable data results in narrower intervals, as observed in Australia.

\FloatBarrier

\section{Discussion}

In mortality modeling and forecasting, the HUts model introduces a novel methodology by taking advantage of the universal nonlinearity of signatures and applying signature regression to decompose mortality curves. Unlike the HU model, which relies on the covariance function and is sensitive to outliers, the HUts model learns data dependency through signature coefficients using principal component regression or SVD, effectively reducing the impact of outliers. The principal components obtained thus represent the structure and interactions within the data.

The HUts model performs competitively and often surpasses other HU model variants in various scenarios. Notably, the HUts model achieves these results without the need for weighted methods or robust statistics, which are employed by the wHU and HUrob models to improve their performance under specific conditions. This approach of the HUts model delivers robustness and is capable of adapting to diverse datasets. Whether dealing with irregularities as seen in the French mortality data or handling more stable demographic trends like those observed in Australia, the HUts model maintains high accuracy and reliability, making it a highly viable option for actuaries and demographers seeking dependable mortality forecasts without the complexity of additional methodological adjustments.

The HUts model also offers the advantage of easy modification to enhance performance, unlike the HU model's basis expansion. Therefore, suboptimal results are not indicative of inherent limitations but rather reflect its modular nature, allowing for greater adaptability. For instance, the embedding step is not limited to the basepoint time lead-lag transformation, \citet{morrill2021} compiled various other augmentations that can be incorporated to improve the model and better represent the geometric characteristics of a path. Besides augmentations, applying the log-signature \citep{morrill2021} or a randomized signature \citep{Cuchiero2020DiscreteTimeSA,CompagnoniRSL} (see \ref{AppendixG} in the Appendix) over the signature can also be considered to reduce the feature vector dimensions. \citet{morrill2021} also suggests that the scaling of paths or signatures may offer improvements.

It is also important to acknowledge the limitations of this study. The prediction intervals constructed under the HUts model face certain limitations, particularly due to the normality assumptions on error terms. Normality assumptions, when not justified, can lead to inaccuracies in the predicted intervals, necessitating the use of bootstrap methods. Bootstrap methods, although useful in addressing non-normality, come with more computational demand. This computational demand can be a significant drawback in practical applications where computational efficiency is crucial. In addition, bootstrap methods are sensitive to outliers which can affect the resampling process, leading to wider prediction intervals. This is evident in the case of French mortality data, which includes periods of war and disease outbreaks. These outliers introduce greater variability into the bootstrap samples, resulting in more conservative and wider intervals. This phenomenon is less pronounced in the mortality data, where the historical data is more stable and follows a general trend without significant outliers. Steps such as adapting the robust bootstrap procedure from \citet{beyshang2022} can be taken to reduce the influence of outliers by assigning weights to model residuals.

In addition, the performance of the HUts model, like any other model, depends on the country's data and forecast horizon, with some models performing relatively better in certain settings. This emphasizes the influence of country-specific mortality dynamics and model sensitivities. As \citet{Shang2012} pointed out, there is no universal mortality model capable of effectively handling data from all countries. This makes the HUts model a candidate for model averaging approaches. Model averaging combines the strengths of various models to improve forecast accuracy by assigning empirical weights to each model based on their performance at different horizons. Given the HUts model's demonstrated efficacy, especially in long-term forecasts, it can be assigned greater weight in model-averaged forecasts to leverage its strengths \citep{shangbooth2020,Chang_Shi_2023}. The benefits of model averaging lie in enhancing forecast accuracy through a strategic combination of models. By integrating the HUts model into a model averaging framework, it is possible to achieve more reliable and accurate mortality forecasts, particularly over longer forecast horizons.

Beyond demography, the HUts model extends itself by offering a robust framework to address a range of challenges in functional data analysis, including functional principal component regression (FPCR), functional regression with functional responses, and forecasting functional time series. The flexibility of the HUts model to adapt to various functional data analysis scenarios is primarily due to its ability to integrate advanced mathematical concepts from functional data analysis, such as smoothing and handling of high-dimensional datasets. This integration not only enhances the model's performance in demographic forecasting but also makes it a powerful tool for broader applications in statistical and machine learning fields involving complex, structured data.

\section{Conclusion}
This paper has introduced the incorporation of signatures into the HU model, resulting in the HUts model. It is able to outperform the HU model variants while exhibiting enhanced robustness, particularly in the presence of outliers and irregularities in the data. The empirical results from the French and other mortality datasets provide compelling evidence of the HUts model's efficacy. In the case of France, with the presence of notable historical outliers due to wars and disease outbreaks, the HUts model showed substantial improvement over the HU model variants. This is evident from the lower MSE values across various forecast horizons. The enhanced performance can be attributed to the model's ability to mitigate the influence of outliers, which are more prevalent in the French data. This robustness is crucial for accurate mortality forecasting in regions with volatile historical trends. The HUts model's superior performance across all age groups, particularly at longer forecast horizons, displays the model's versatility across diverse demographic conditions. Prediction intervals were constructed with the application of bootstrap procedures. The coverage probabilities of the 95\% bootstrapped prediction intervals generally are close to the intended nominal coverage, while the adjustment is able to improve said coverage for certain cases. The presence of outliers can impact the construction of the intervals, which is particularly pronounced in the case of French mortality, where the intervals constructed were more conservative and wide. In contrast, Australian mortality, with fewer outliers, shows smaller coverage deviance with its bootstrapped prediction interval. From a practical standpoint, these findings have significant implications for actuaries and policymakers. The improved accuracy and robustness of the HUts model can lead to more reliable mortality forecasts, which are essential for various applications, including insurance pricing, pension fund management, and public health planning. The model's ability to handle data with outliers and provide accurate long-term forecasts makes it a valuable tool for demography.

Future research should focus on refining the HUts model, potentially integrating more advanced smoothing techniques or exploring alternative statistical methods to enhance its performance further. Additionally, the model's application can be extended to multi-population mortality forecasting as signatures are an excellent dimension reduction tool, and incorporating socioeconomic factors as a binary dimension could provide deeper insights into mortality trends. The following practical extensions can be readily implemented:

\begin{enumerate}
    \item Gender specific HUts model
        \begin{align}
        \text{Male: } f_{t,M}(x)=\mu(x)+\sum_{k=1}^K \beta_{t,k} Z_{k}(x) +e_{t}(x),\nonumber \\
        \text{Female: } f_{t,F}(x)=\mu(x)+\sum_{k=1}^K \beta_{t,k} Z_{k}(x) +e_{t}(x).\nonumber
        \end{align}
    This adaptation allows the model to capture and forecast mortality trends separately for different genders, taking into account potential differences in mortality patterns, improvements, and other gender-specific characteristics.
    \item Fitting the HUts model with mortality improvement rates, $z_{x,t}=2\times\frac{1-\frac{m_{x,t}}{m_{x,t-1}}}{1+\frac{m_{x,t}}{m_{x,t-1}}}$ instead of raw mortality rates $m_{x,t}$ \citep{HABERMAN2012309}. Let the underlying smoothed function of $z_{x,t}$ be $\gamma_{t}(x)$, then using the HUts we can decompose it into:
    \begin{equation}
        \gamma_{t}(x)=\mu(x)+\sum_{k=1}^K \beta_{t,k} Z_{k}(x) +e_{t}(x). \nonumber
    \end{equation}
    Modeling mortality improvement rates helps stabilize the mortality data series by focusing on changes over time, making it stationary and enhancing the accuracy of forecasts.
    \item Coherent mortality forecasting using the product ratio method by \citet{hyn2013} by fitting $p_t(x) = \sqrt{f_{t,M}(x) \cdot f_{t,F}(x)}$ and
    $r_t(x) = \sqrt{\frac{f_{t,M}(x)}{f_{t,F}(x)}}$ using the HUts model and forecasting it $h$-steps-ahead. 
    Then the $h$-steps-ahead forecasts of the gender specific log mortality rates can be obtained as follows:
        \begin{align}
         f_{n+h|n,M}(x)&=p_{n+h|n}(x) \times r_{n+h|n}(x), \nonumber\\
          f_{n+h|n,F}(x)&=\frac{p_{n+h|n}(x)}{r_{n+h|n}(x)}.\nonumber  
        \end{align}
    This method ensures that the mortality forecasts for different subpopulations remain within realistic and historically observed relationships. \citet{cristian2024,SHANG2022239,shang2017} also provide potential directions for extending the HUts model for subpopulation modelling.
\end{enumerate}

\bibliographystyle{plainnat}
\bibliography{library.bib}

\begin{appendix}
\section{Appendix}
    \subsection{Normality assumptions of errors}\label{AppendixA}
\begin{figure}[htbp!]
    \centering

    \begin{subfigure}[b]{0.45\linewidth} 
        \centering
        \includegraphics[width=\linewidth]{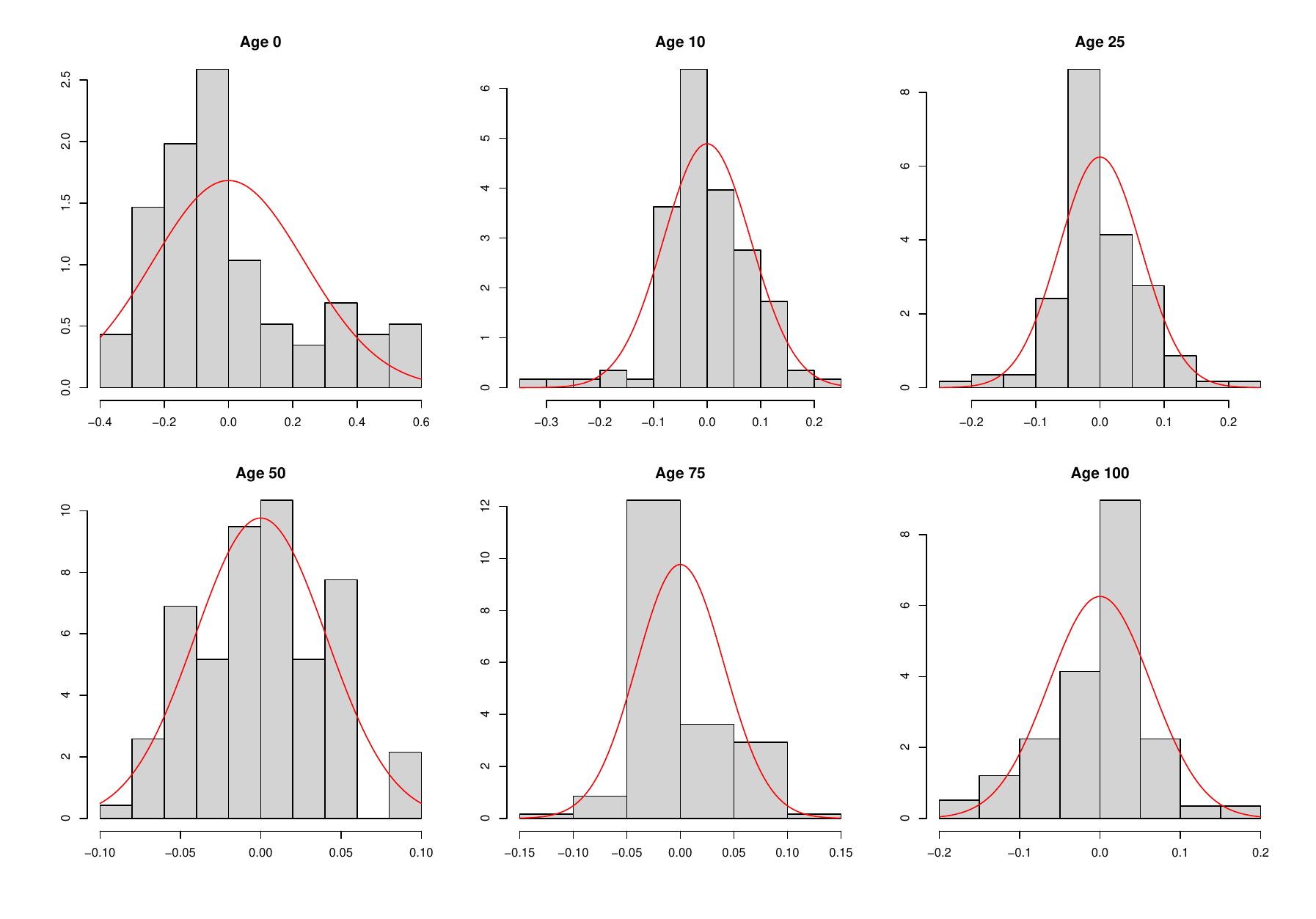}
        \caption{Histogram}
        \label{fig:normHUts}
    \end{subfigure}
    \hfill
    \begin{subfigure}[b]{0.45\linewidth} 
        \centering
        \includegraphics[width=\linewidth]{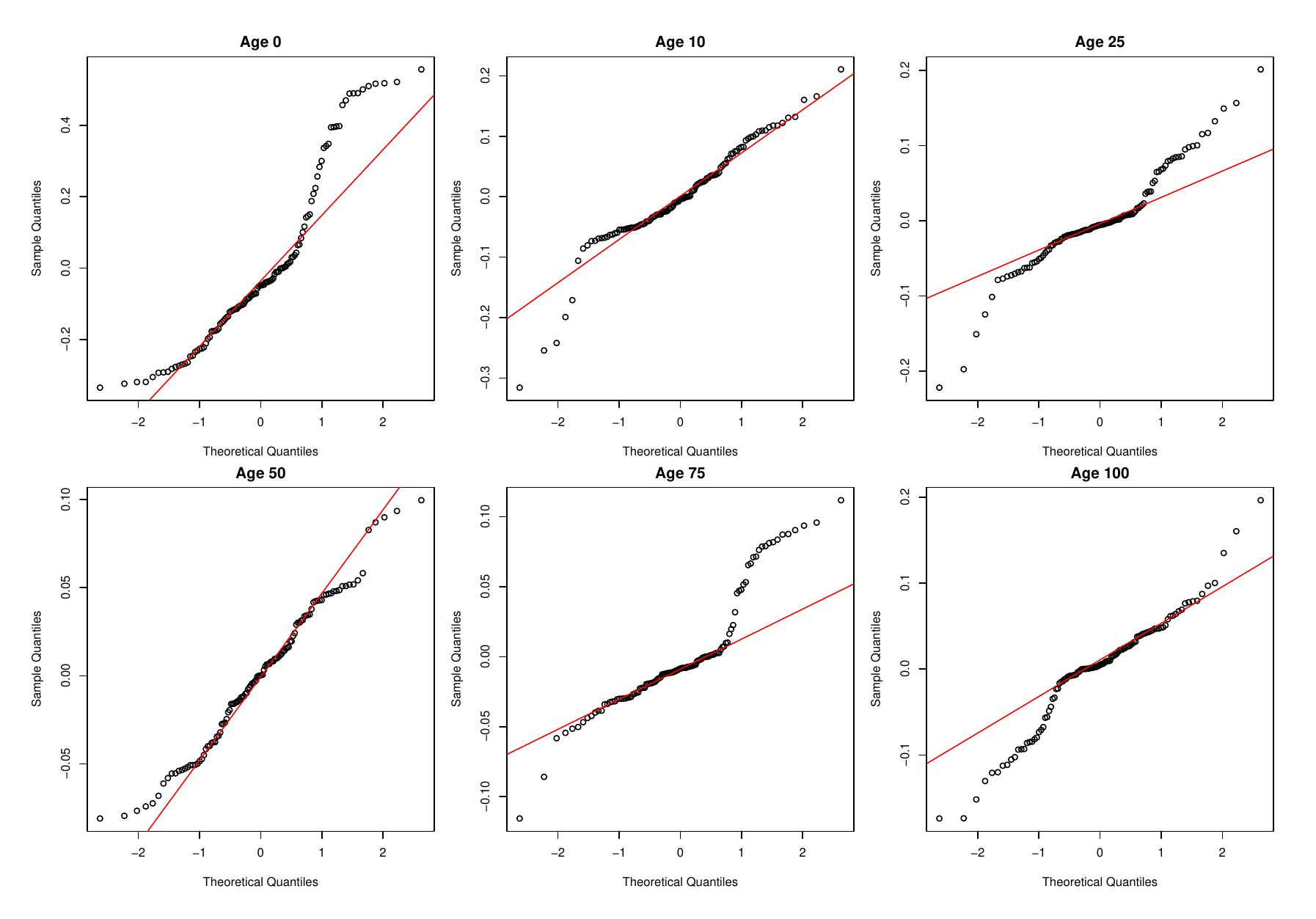} 
        \caption{QQ-plot}
        \label{fig:qqhuts}
    \end{subfigure}
    \caption{HUts model residuals}
    \label{fig:HUtsresid}
\end{figure}

\begin{figure}[htbp!]
    \centering

    \begin{subfigure}[b]{0.45\linewidth} 
        \centering
        \includegraphics[width=\linewidth]{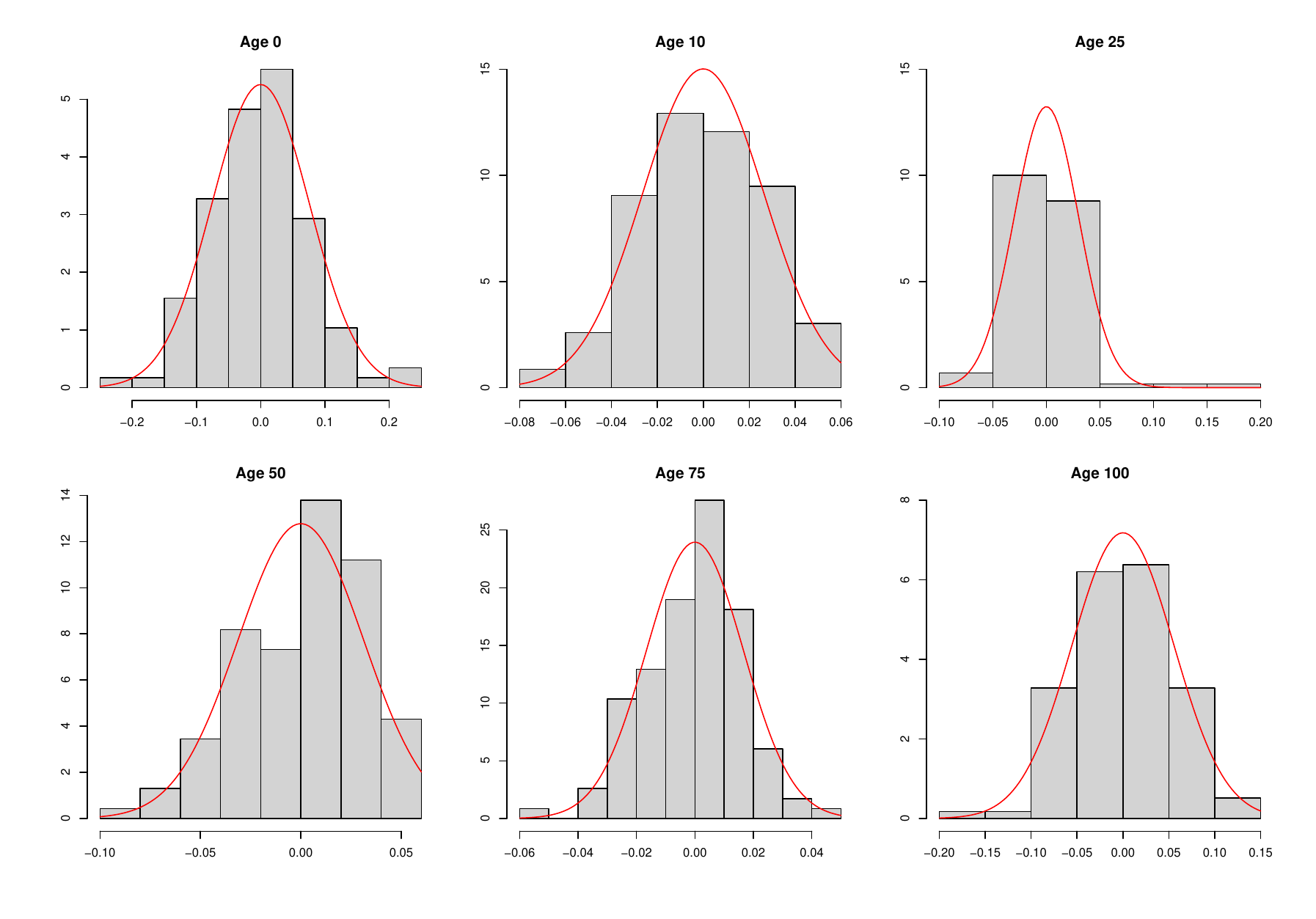}
        \caption{Histogram}
        \label{fig:normHU}
    \end{subfigure}
    \hfill
    \begin{subfigure}[b]{0.45\linewidth} 
        \centering
        \includegraphics[width=\linewidth]{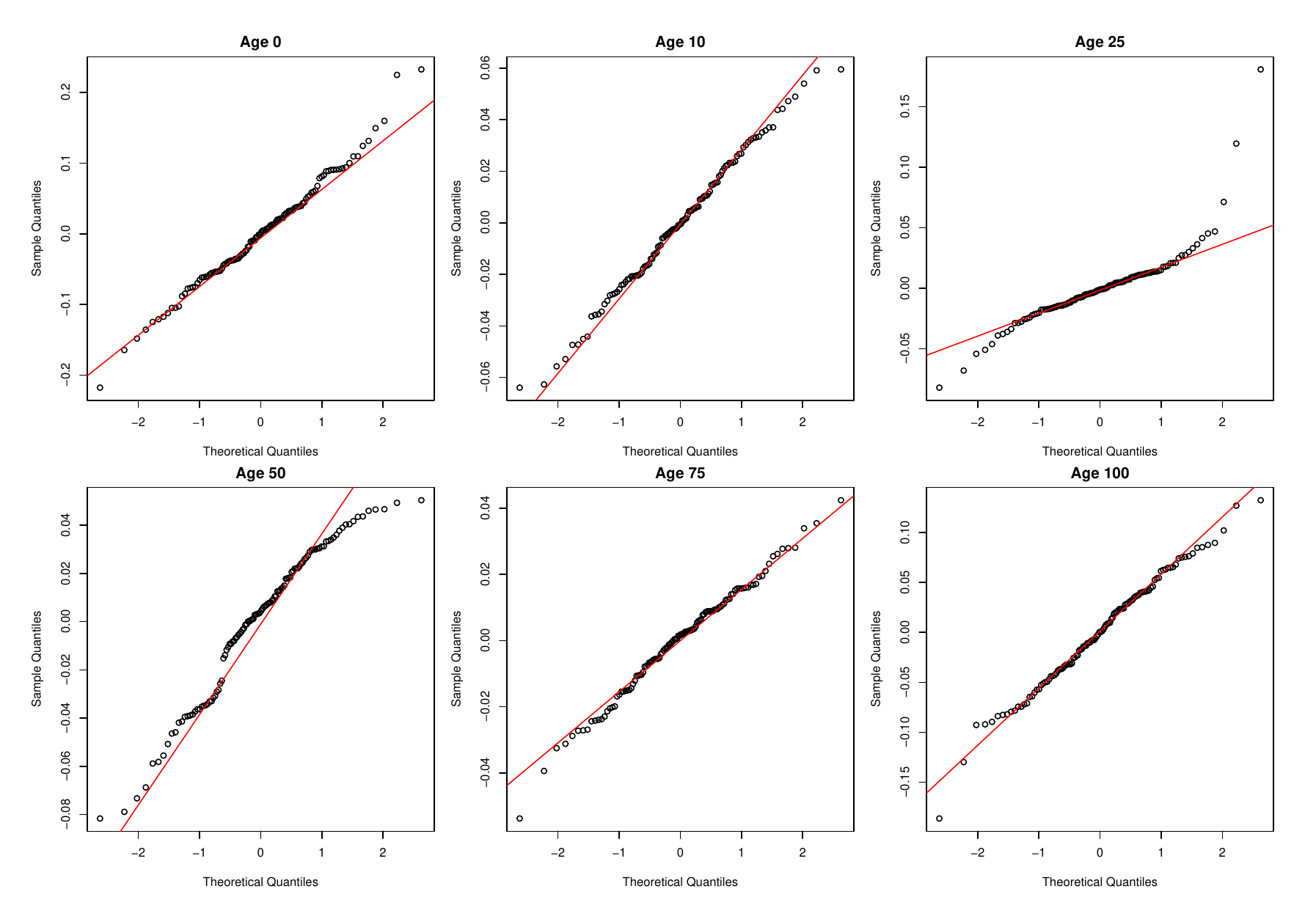} 
        \caption{QQ-plot}
        \label{fig:qqhu}
    \end{subfigure}
    \caption{HU model residuals}
    \label{fig:HUresid}
\end{figure}

\begin{figure}[htbp!]
    \centering

    \begin{subfigure}[b]{0.45\linewidth} 
        \centering
        \includegraphics[width=\linewidth]{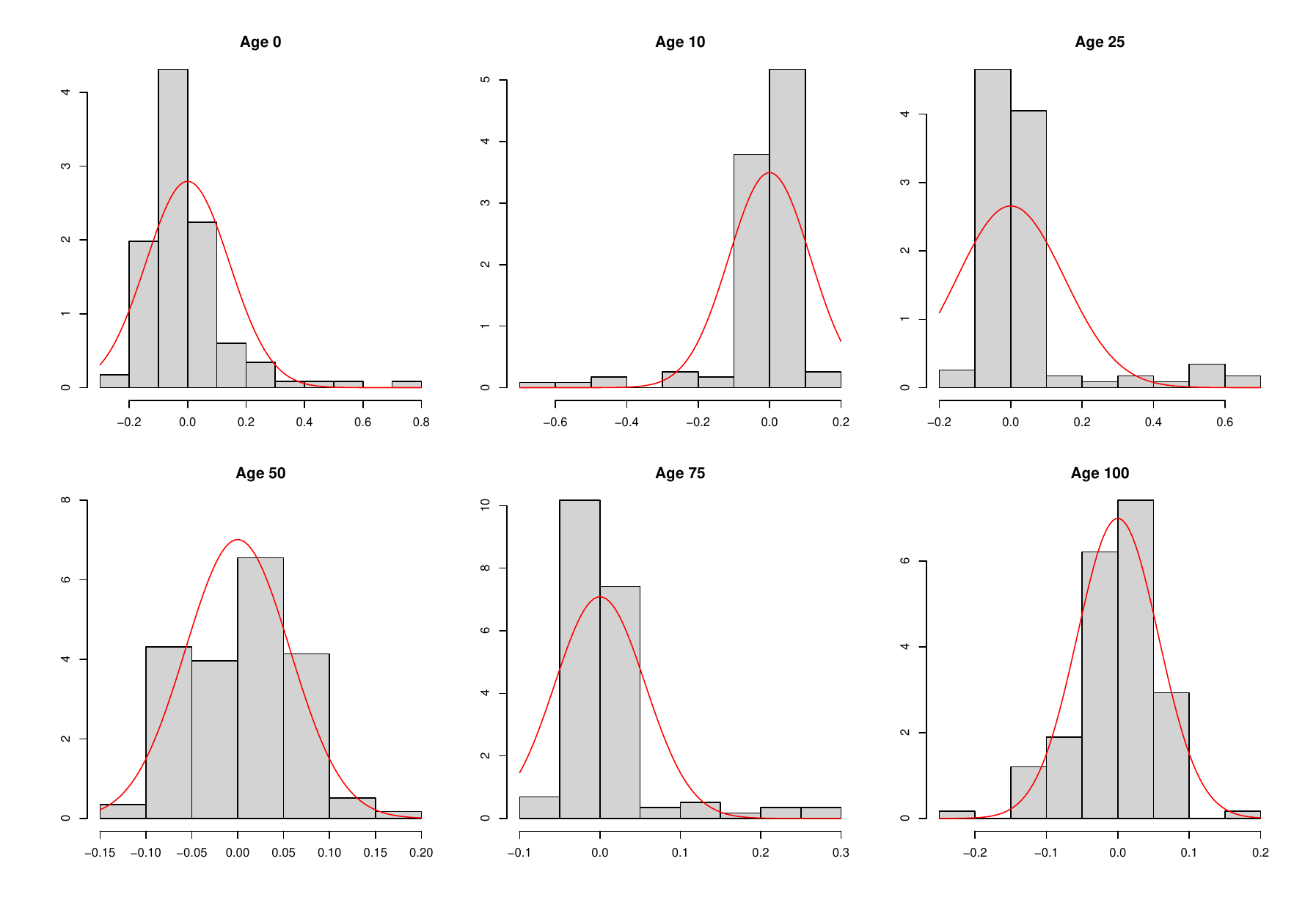}
        \caption{Histogram}
        \label{fig:normHUrob}
    \end{subfigure}
    \hfill
    \begin{subfigure}[b]{0.45\linewidth} 
        \centering
        \includegraphics[width=\linewidth]{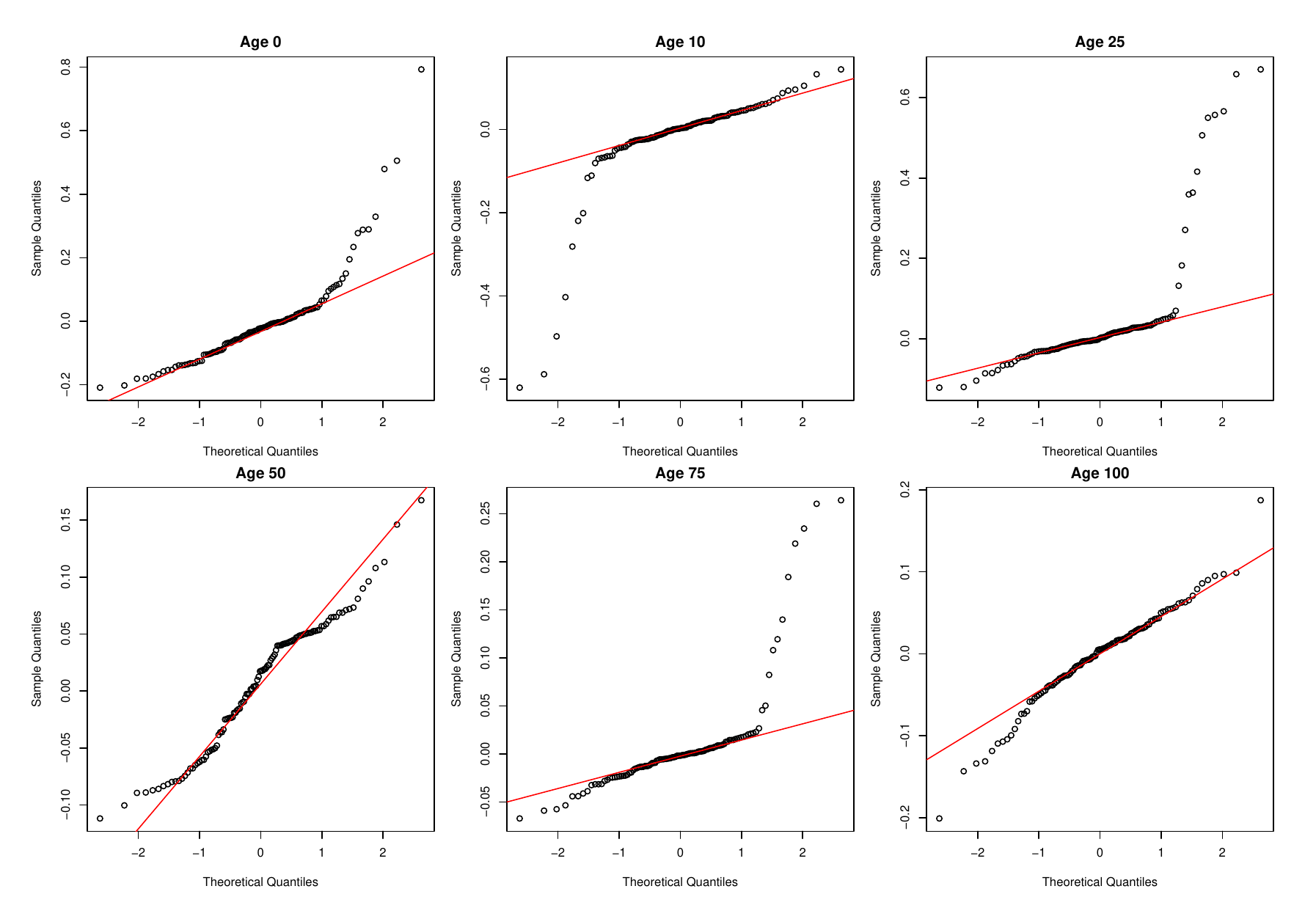} 
        \caption{QQ-plot}
        \label{fig:qqhurob}
    \end{subfigure}
    \caption{HUrob model residuals}
    \label{fig:HUrobresid}
\end{figure}

\begin{figure}[htbp!]
    \centering

    \begin{subfigure}[b]{0.45\linewidth} 
        \centering
        \includegraphics[width=\linewidth]{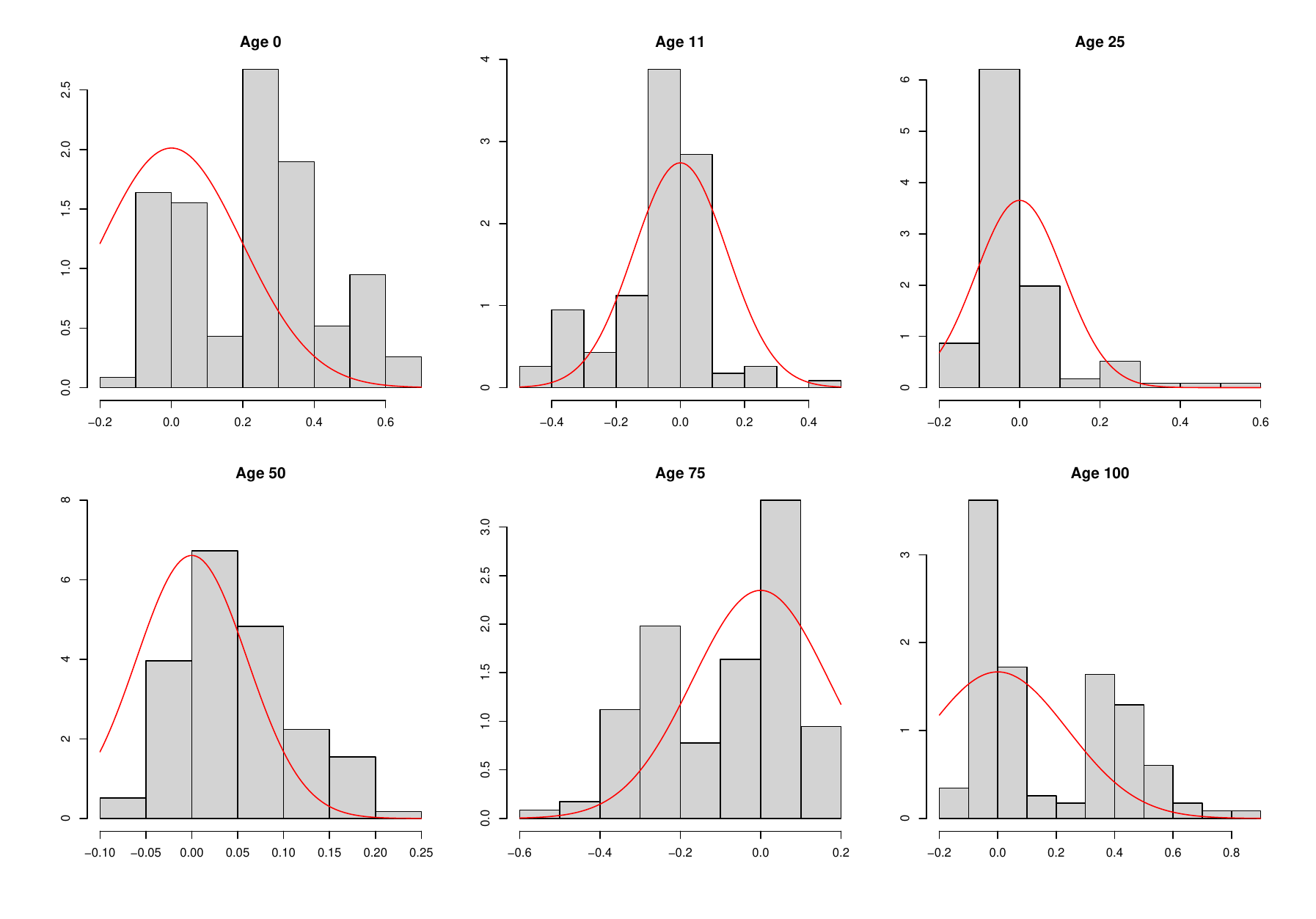}
        \caption{Histogram}
        \label{fig:normwHU}
    \end{subfigure}
    \hfill
    \begin{subfigure}[b]{0.45\linewidth} 
        \centering
        \includegraphics[width=\linewidth]{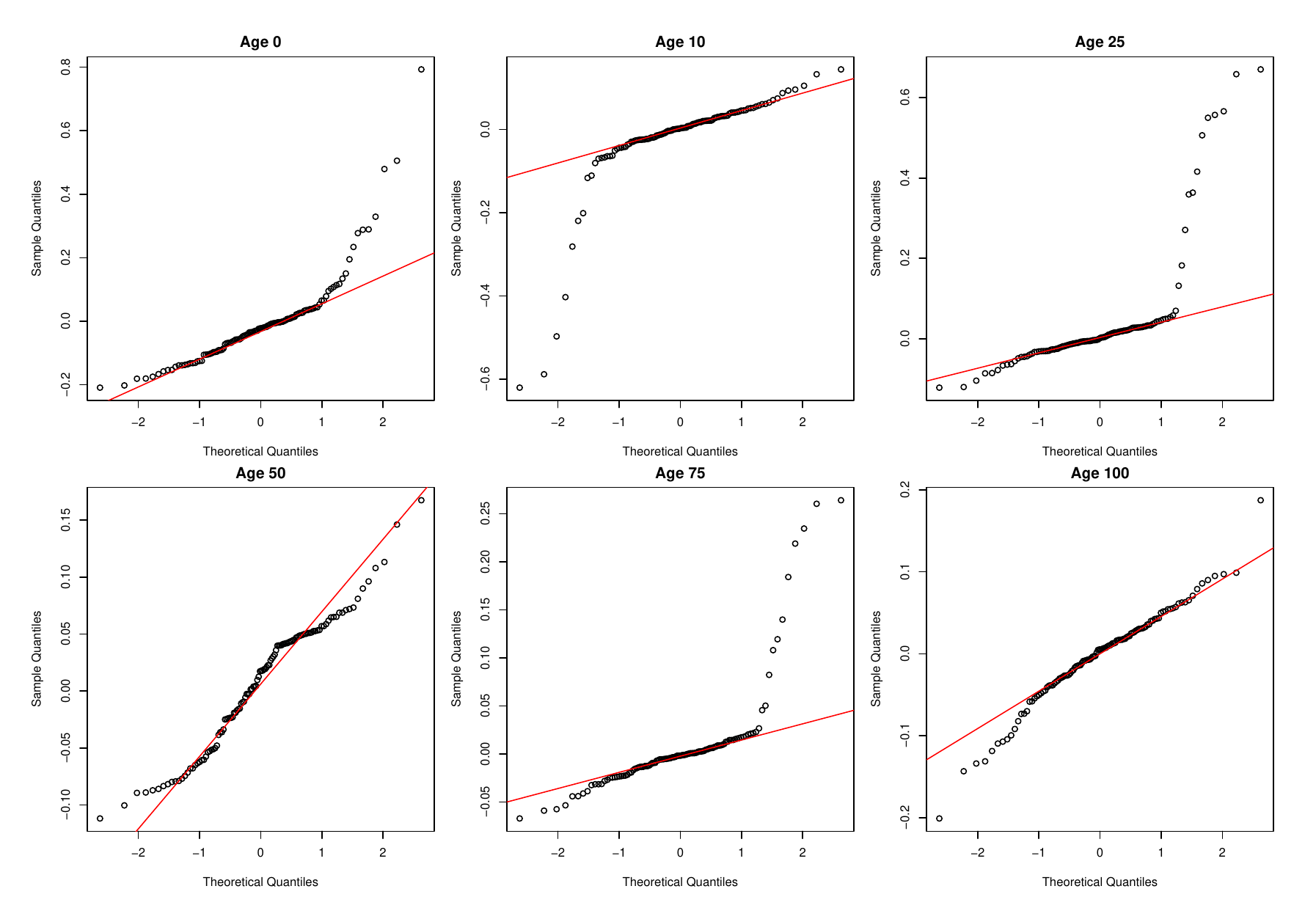} 
        \caption{QQ-plot}
        \label{fig:qqwhu}
    \end{subfigure}
    \caption{wHU model residuals}
    \label{fig:wHUresid}
\end{figure}

To obtain prediction intervals using distributional forecasts, all sources of error must satisfy the normality assumptions. If any source of error does not follow a normal distribution, bootstrapping methods should be used instead. While it is unrealistic to expect perfect normality, we prefer that most of the residuals resemble a normal distribution; otherwise, bootstrapped methods are preferred. We have plotted the residual histograms and QQ plots of ages 0, 10, 25, 50, 75, and 100 for all four models fitted with French (1899-2014) mortality data . Only the residuals of the HU model, seen in Figure \ref{fig:HUresid}, seem to resemble somewhat of a normal distribution. The residuals from the three remaining models, seen in Figures \ref{fig:HUtsresid}, \ref{fig:HUrobresid}, and \ref{fig:wHUresid}, deviate significantly from a normal distribution and would benefit from the bootstrapping procedure when constructing prediction intervals.

\FloatBarrier
\subsection{Selection of truncation order}\label{AppendixB}

\begin{figure}[htbp!]
\centering
\includegraphics[width=0.75\linewidth]{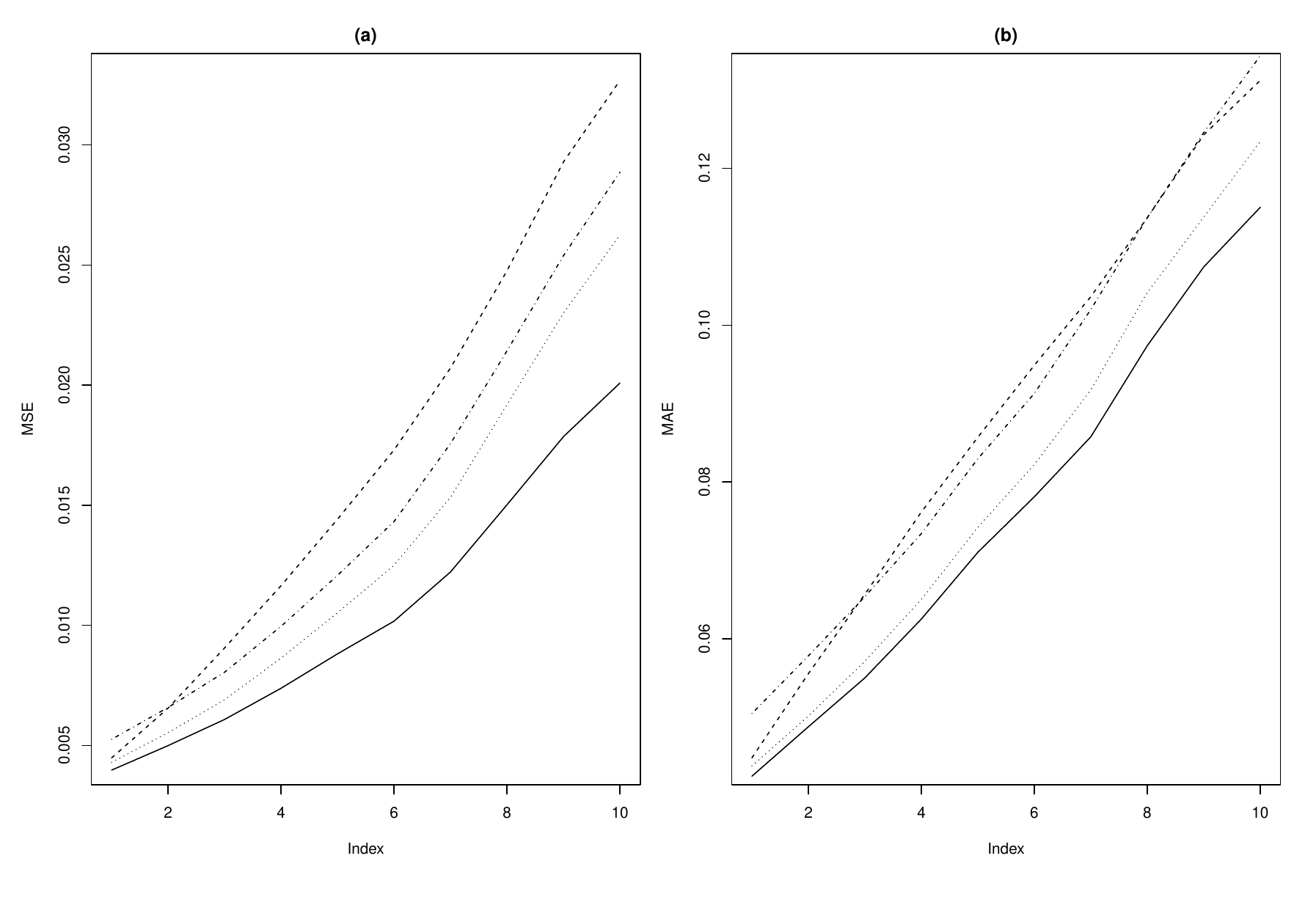}
\caption{(a) MSE and (b) MAE plots for French mortality of the HUts model with $m=2$ (solid line), $m=3$ (dashed line), $m=4$ (dotted line), and $m=5$ (dash-dotted line)}
\label{fig: trunc}
\end{figure}

As the number of signature coefficients increases exponentially with the truncation order, we employ a naive search based on the MSE and MAE of the model's forecasting performance. In signature regression, it is generally advisable to keep the truncation order, $m$, below 5. A truncation order of 5 already includes 182 signature coefficients, significantly impacting the computational effort required. This approach is demonstrated using French mortality data. As depicted in Figure \ref{fig: trunc}, the MSE and MAE values of the HUts model with $m=3,4,5$ are larger than when $m=2$. Since our path dimension is only 3, a high order of signature coefficients may not be necessary to accurately represent the path.

\subsection{Data availability and codes}\label{AppendixC}
The codes are available upon request. 

\subsection{Mean squared error and mean absolute error tables}\label{AppendixD}
\begin{table}[htbp]
\centering
\caption{MSE of one-step-ahead point forecasts of log mortality rates by method and country}
\label{tab:MSE1}
\begin{tabular}{lcccc}
\toprule
& HUts & HU & HUrob & wHU \\
\midrule
Australia     & 0.00982 & 0.01027 & 0.01696 & \textbf{0.00952} \\
Belgium       & 0.02183 & 0.02144 & 0.02419 & \textbf{0.02047} \\
Bulgaria      & \textbf{0.01942} & 0.02131 & 0.02158 & 0.01968 \\
Denmark       & 0.03636 & 0.03810 & 0.04098 & \textbf{0.03590} \\
Finland       & 0.03856 & 0.04271 & 0.04435 & \textbf{0.03705} \\
France        & \textbf{0.00398} & 0.00642 & 0.00710 & 0.00399 \\
Ireland       & \textbf{0.05062} & 0.05174 & 0.05735 & 0.05161 \\
Italy         & 0.00711 & 0.00920 & 0.01600 & \textbf{0.00680} \\
Japan         & \textbf{0.00674} & 0.00747 & 0.00938 & 0.00687 \\
Netherlands   & 0.01242 & 0.01344 & 0.01569 & \textbf{0.01143} \\
Norway        & 0.04397 & 0.04303 & 0.04800 & \textbf{0.04264} \\
United States & 0.00133 & 0.00207 & 0.00535 & \textbf{0.00129} \\
\midrule
Mean          & 0.02101 & 0.02227 & 0.02558 & \textbf{0.02061} \\
\bottomrule
\end{tabular}
\end{table}

\begin{table}[htbp]
\centering
\caption{MSE of five-step-ahead point forecasts of log mortality rates by method and country}
\label{tab:MSE5}
\begin{tabular}{lcccc}
\toprule
& HUts & HU & HUrob & wHU \\
\midrule
Australia     & 0.01835 & 0.02245 & 0.02890 & \textbf{0.01642} \\
Belgium       & 0.02627 & 0.02863 & 0.03211 & \textbf{0.02620} \\
Bulgaria      & 0.04595 & 0.05212 & \textbf{0.04574} & 0.05478 \\
Denmark       & \textbf{0.04891} & 0.05484 & 0.05764 & 0.05007 \\
Finland       & 0.04698 & 0.05988 & 0.05328 & \textbf{0.04538} \\
France        & \textbf{0.00882} & 0.02290 & 0.01489 & 0.00978 \\
Ireland       & \textbf{0.07035} & 0.07198 & 0.07707 & 0.07785 \\
Italy         & 0.01625 & 0.03153 & 0.02982 & \textbf{0.01546} \\
Japan         & 0.01045 & 0.01288 & 0.01361 & \textbf{0.01020} \\
Netherlands   & \textbf{0.01785} & 0.02171 & 0.02374 & 0.01859 \\
Norway        & 0.05056 & 0.05309 & 0.05856 & \textbf{0.05053} \\
United States & \textbf{0.00746} & 0.00900 & 0.01252 & 0.00896 \\
\midrule
Mean          & \textbf{0.03068} & 0.03675 & 0.03732 & 0.03202 \\
\bottomrule
\end{tabular}
\end{table}

\begin{table}[htbp]
\centering
\caption{MSE of ten-step-ahead point forecasts of log mortality rates by method and country}
\label{tab:MSE10}
\begin{tabular}{lcccc}
\toprule
& HUts & HU & HUrob & wHU \\
\midrule
Australia     & 0.0418  & 0.0511  & 0.0509  & \textbf{0.0351} \\
Belgium       & \textbf{0.0364}  & 0.0380  & 0.0450  & 0.0404 \\
Bulgaria      & 0.0976  & 0.0931  & \textbf{0.0811}  & 0.1280 \\
Denmark       & \textbf{0.0764}  & 0.0825  & 0.0927  & 0.0825 \\
Finland       & 0.0653  & 0.0894  & 0.0773  & \textbf{0.0617} \\
France        & \textbf{0.0201}  & 0.0526  & 0.0278  & 0.0242 \\
Ireland       & \textbf{0.1111}  & 0.1297  & 0.1177  & 0.1413 \\
Italy         & \textbf{0.0403}  & 0.0603  & 0.0594  & 0.0433 \\
Japan         & 0.0348  & 0.0399  & \textbf{0.0289}  & 0.0362 \\
Netherlands   & \textbf{0.0326}  & 0.0391  & 0.0406  & 0.0377 \\
Norway        & \textbf{0.0695}  & 0.0775  & 0.0802  & 0.0718 \\
United States & 0.0226  & \textbf{0.0196}  & 0.0342  & 0.0305 \\
\midrule
Mean          & \textbf{0.0540}  & 0.0644  & 0.0613  & 0.0611 \\
\bottomrule
\end{tabular}
\end{table}

\FloatBarrier

The mean absolute error expresses the average difference between predicted and observed values and is computed using:
\begin{equation}
    MAE(h) = \frac{1}{pq}\sum^q_{t=1}\sum^p_{i=1} \lvert y_t(x_i)-\hat{y}_{t|t-h}(x_i) \rvert.  \nonumber
\end{equation}

\begin{table}[htbp]
\centering
\caption{MAE of one-step-ahead point forecasts of log mortality rates by method and country}
\label{tab:MAE1}
\begin{tabular}{lcccc}
\toprule
& HUts & HU & HUrob & wHU \\
\midrule
Australia     & 0.06111 & 0.06614 & 0.09498 & \textbf{0.06036} \\
Belgium       & 0.08404 & 0.09092 & 0.10116 & \textbf{0.08193} \\
Bulgaria      & \textbf{0.09106} & 0.10052 & 0.09727 & 0.09138 \\
Denmark       & 0.11169 & 0.12029 & 0.12909 & \textbf{0.11127} \\
Finland       & 0.11279 & 0.12608 & 0.13552 & \textbf{0.10976} \\
France        & \textbf{0.04246} & 0.06100 & 0.06336 & 0.04256 \\
Ireland       & \textbf{0.12512} & 0.13419 & 0.14838 & 0.12719 \\
Italy         & 0.05330 & 0.06842 & 0.08936 & \textbf{0.05210} \\
Japan         & 0.05061 & 0.05701 & 0.06955 & \textbf{0.04975} \\
Netherlands   & 0.06802 & 0.07839 & 0.08692 & \textbf{0.06657} \\
Norway        & 0.11718 & 0.11890 & 0.13299 & \textbf{0.11479} \\
United States & 0.02648 & 0.03459 & 0.05380 & \textbf{0.02628} \\
\midrule
Mean          & 0.07866 & 0.08804 & 0.10020 & \textbf{0.07783} \\
\bottomrule
\end{tabular}
\end{table}

\begin{table}[htbp]
\centering
\caption{MAE of five-step-ahead point forecasts of log mortality rates by method and country}
\label{tab:MAE5}
\begin{tabular}{lcccc}
\toprule
& HUts & HU & HUrob & wHU \\
\midrule
Australia     & 0.09131 & 0.11006 & 0.12925 & \textbf{0.08682} \\
Belgium       & \textbf{0.10163} & 0.11696 & 0.12482 & 0.10307 \\
Bulgaria      & 0.15511 & 0.17190 & \textbf{0.15103} & 0.15636 \\
Denmark       & \textbf{0.14989} & 0.16413 & 0.17112 & 0.15113 \\
Finland       & 0.14040 & 0.17654 & 0.15844 & \textbf{0.13800} \\
France        & \textbf{0.07102} & 0.12234 & 0.09527 & 0.07427 \\
Ireland       & \textbf{0.16031} & 0.18212 & 0.18869 & 0.17752 \\
Italy         & 0.09528 & 0.14200 & 0.13301 & \textbf{0.08831} \\
Japan         & 0.07659 & 0.08745 & 0.08784 & \textbf{0.07277} \\
Netherlands   & \textbf{0.09511} & 0.11317 & 0.11668 & 0.09702 \\
Norway        & 0.13801 & 0.15506 & 0.16166 & \textbf{0.13688} \\
United States & \textbf{0.06428} & 0.07333 & 0.08508 & 0.06843 \\
\midrule
Mean          & \textbf{0.11158} & 0.13459 & 0.13357 & 0.11255 \\
\bottomrule
\end{tabular}
\end{table}

\begin{table}[htbp]
\centering
\caption{MAE of ten-step-ahead point forecasts of log mortality rates by method and country}
\label{tab:MAE10}
\begin{tabular}{lcccc}
\toprule
& HUts & HU & HUrob & wHU \\
\midrule
Australia     & 0.1483  & 0.1705  & 0.1749  & \textbf{0.1383} \\
Belgium       & \textbf{0.1323}  & 0.1421  & 0.1558  & 0.1434 \\
Bulgaria      & 0.2490  & 0.2389  & \textbf{0.2093}  & 0.2341 \\
Denmark       & \textbf{0.2097}  & 0.2182  & 0.2352  & 0.2151 \\
Finland       & 0.1804  & 0.2334  & 0.2105  & \textbf{0.1754} \\
France        & \textbf{0.1150}  & 0.1905  & 0.1360  & 0.1204 \\
Ireland       & \textbf{0.2164}  & 0.2663  & 0.2510  & 0.2641 \\
Italy         & 0.1623  & 0.2056  & 0.2019  & \textbf{0.1550} \\
Japan         & 0.1410  & 0.1536  & \textbf{0.1289}  & 0.1435 \\
Netherlands   & \textbf{0.1448}  & 0.1601  & 0.1637  & 0.1564 \\
Norway        & \textbf{0.1809}  & 0.2089  & 0.2087  & 0.1837 \\
United States & \textbf{0.1121}  & 0.1122  & 0.1380  & 0.1250 \\
\midrule
Mean          & \textbf{0.1660}  & 0.1917  & 0.1845  & 0.1712 \\
\bottomrule
\end{tabular}
\end{table}

\FloatBarrier

\FloatBarrier
\subsection{Remaining MSE and MAE plots}\label{AppendixE}
In this section, we categorized the MSE and MAE plots of the remaining 10 countries into two groups: those favorable to the HUts model and those where the HUts model is comparable to the other models.

\subsubsection{Favourable to the HUts model}
\begin{figure}[htbp!]
\centering
\includegraphics[width=0.75\linewidth]{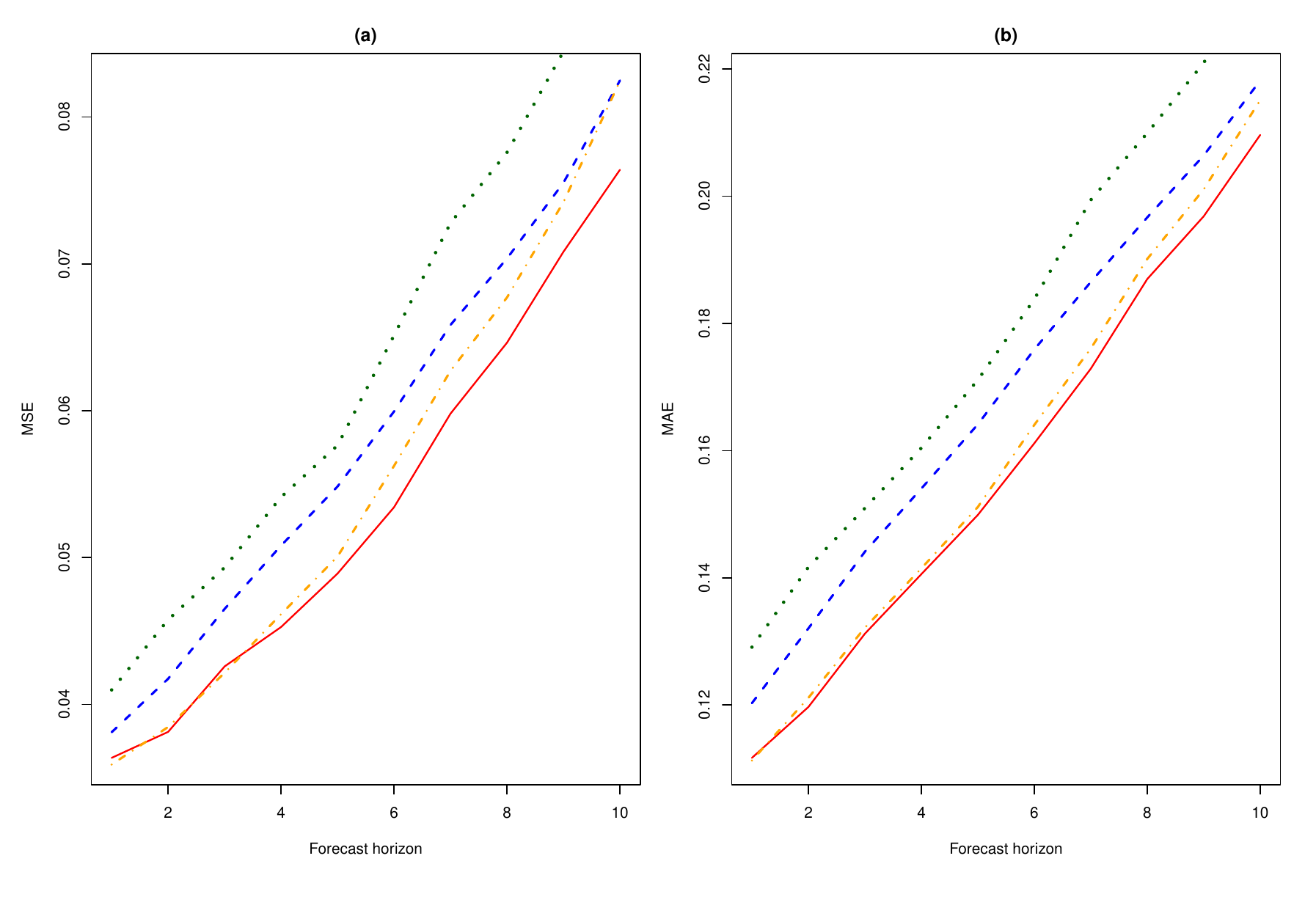}
\caption{(a) MSE and (b) MAE plots for Danish mortality of the HUts (red solid line), HU (blue dashed line), HUrob (green dotted line), and the wHU (orange dash-dotted line) models}
\label{fig: denME}
\end{figure}

\begin{figure}[htbp!]
\centering
\includegraphics[width=0.75\linewidth]{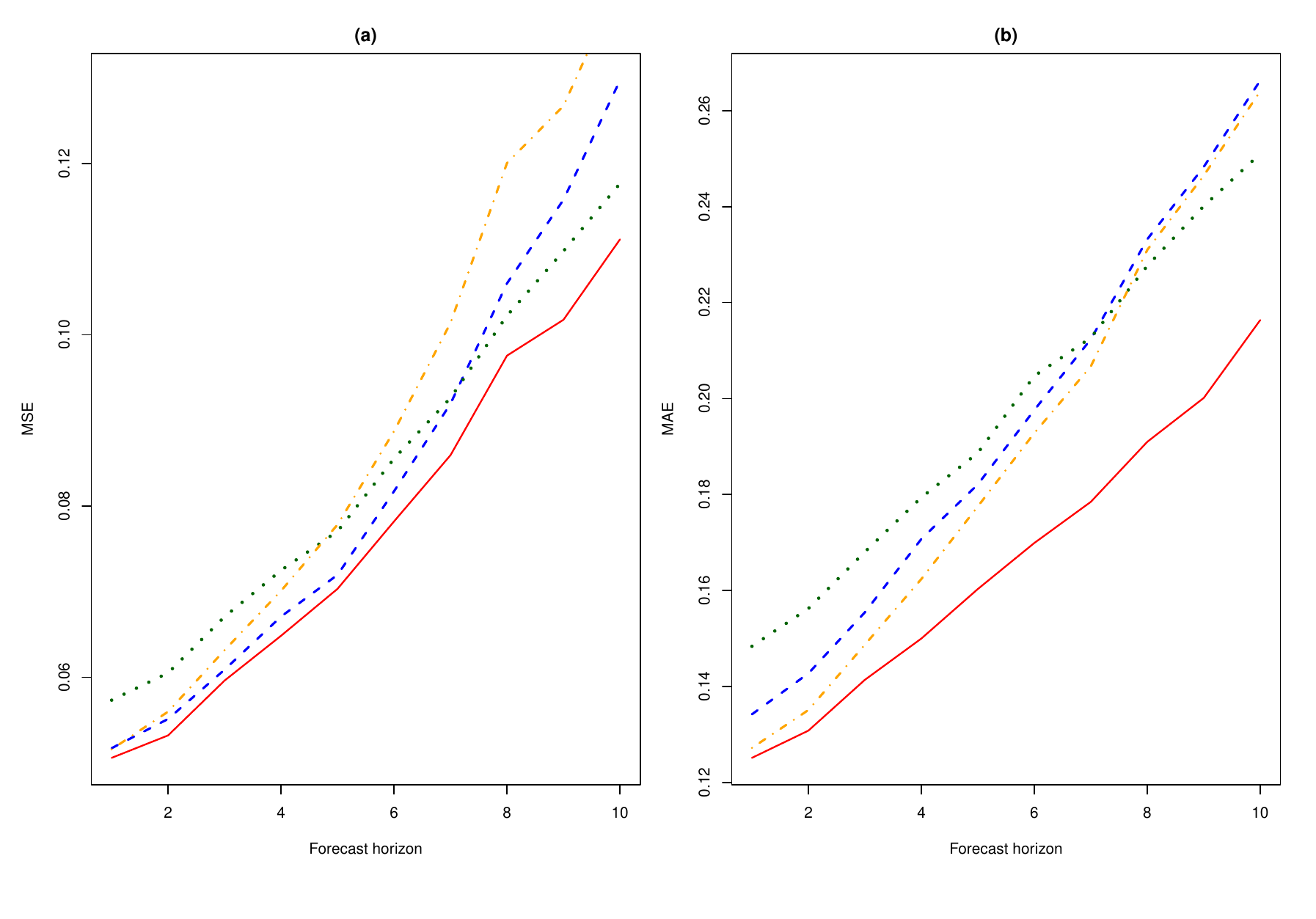}
\caption{(a) MSE and (b) MAE plots for Irish mortality of the HUts (red solid line), HU (blue dashed line), HUrob (green dotted line), and the wHU (orange dash-dotted line) models}
\label{fig: irdME}
\end{figure}

\begin{figure}[htbp!]
\centering
\includegraphics[width=0.75\linewidth]{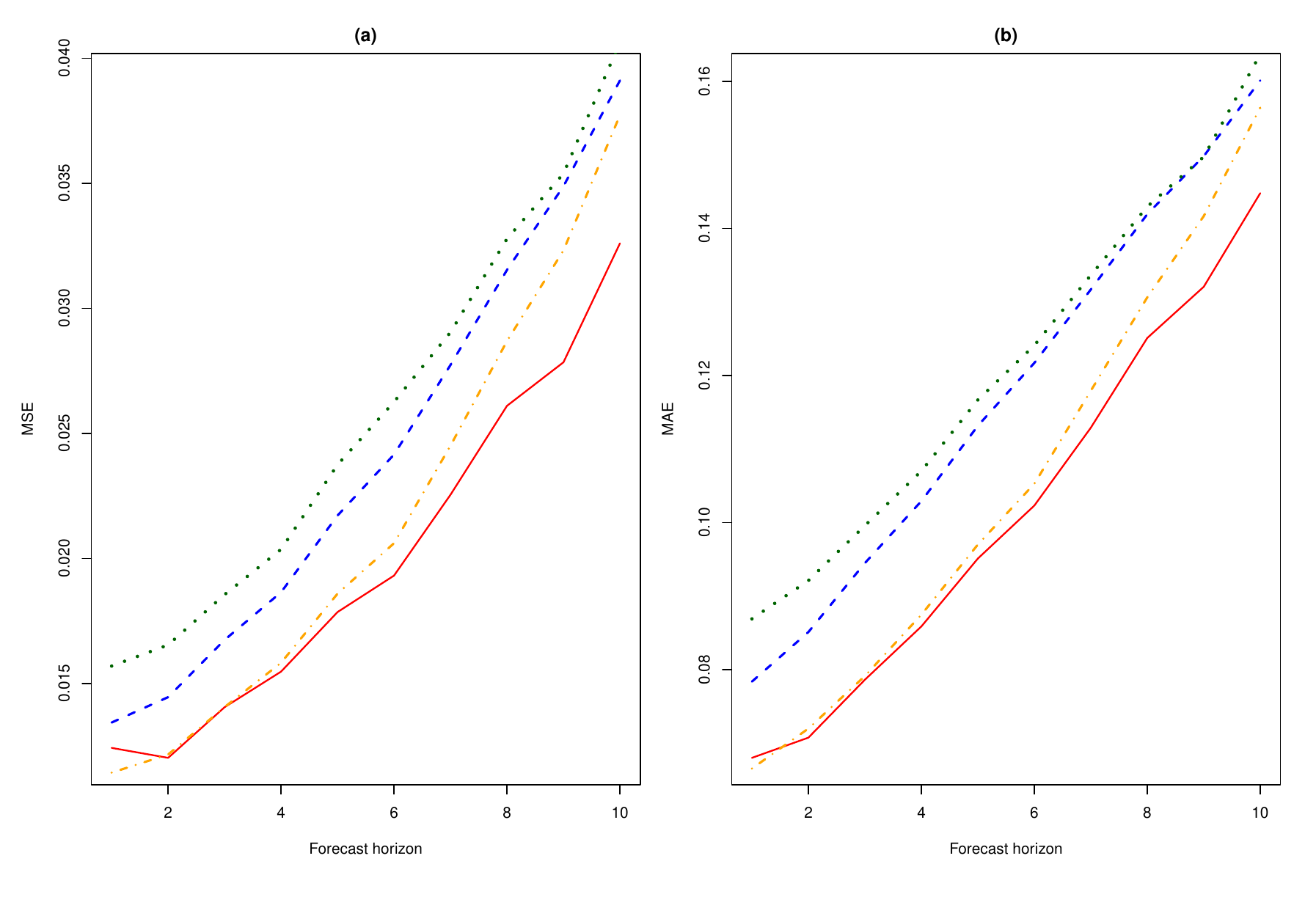}
\caption{(a) MSE and (b) MAE plots for Dutch mortality of the HUts (red solid line), HU (blue dashed line), HUrob (green dotted line), and the wHU (orange dash-dotted line) models}
\label{fig: netME}
\end{figure}

\begin{figure}[htbp!]
\centering
\includegraphics[width=0.75\linewidth]{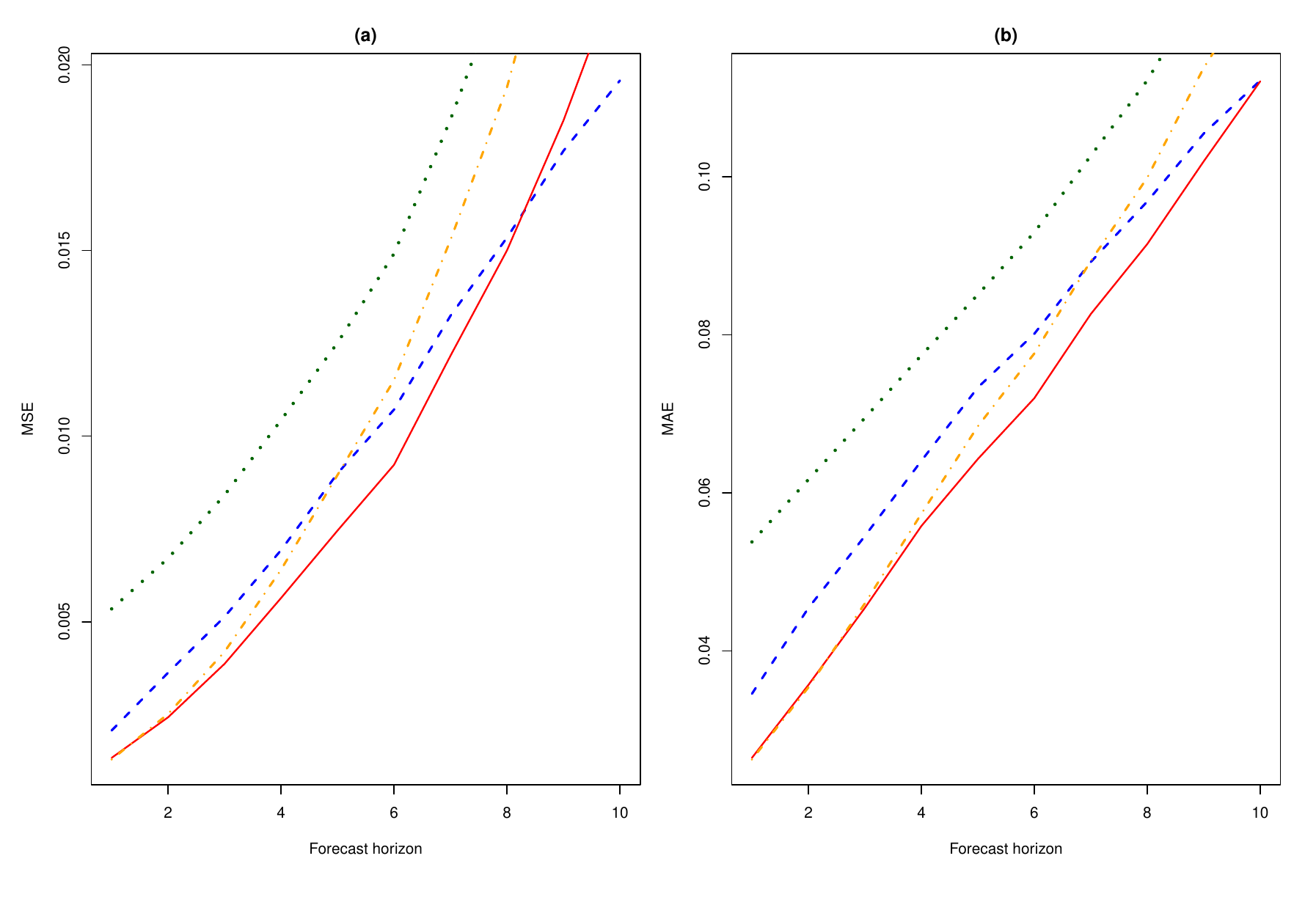}
\caption{(a) MSE and (b) MAE plots for American mortality of the HUts (red solid line), HU (blue dashed line), HUrob (green dotted line), and the wHU (orange dash-dotted line) models}
\label{fig: usaME}
\end{figure}

\FloatBarrier
\clearpage
\subsubsection{Comparable to the HUts model}
\begin{figure}[htbp!]
\centering
\includegraphics[width=0.75\linewidth]{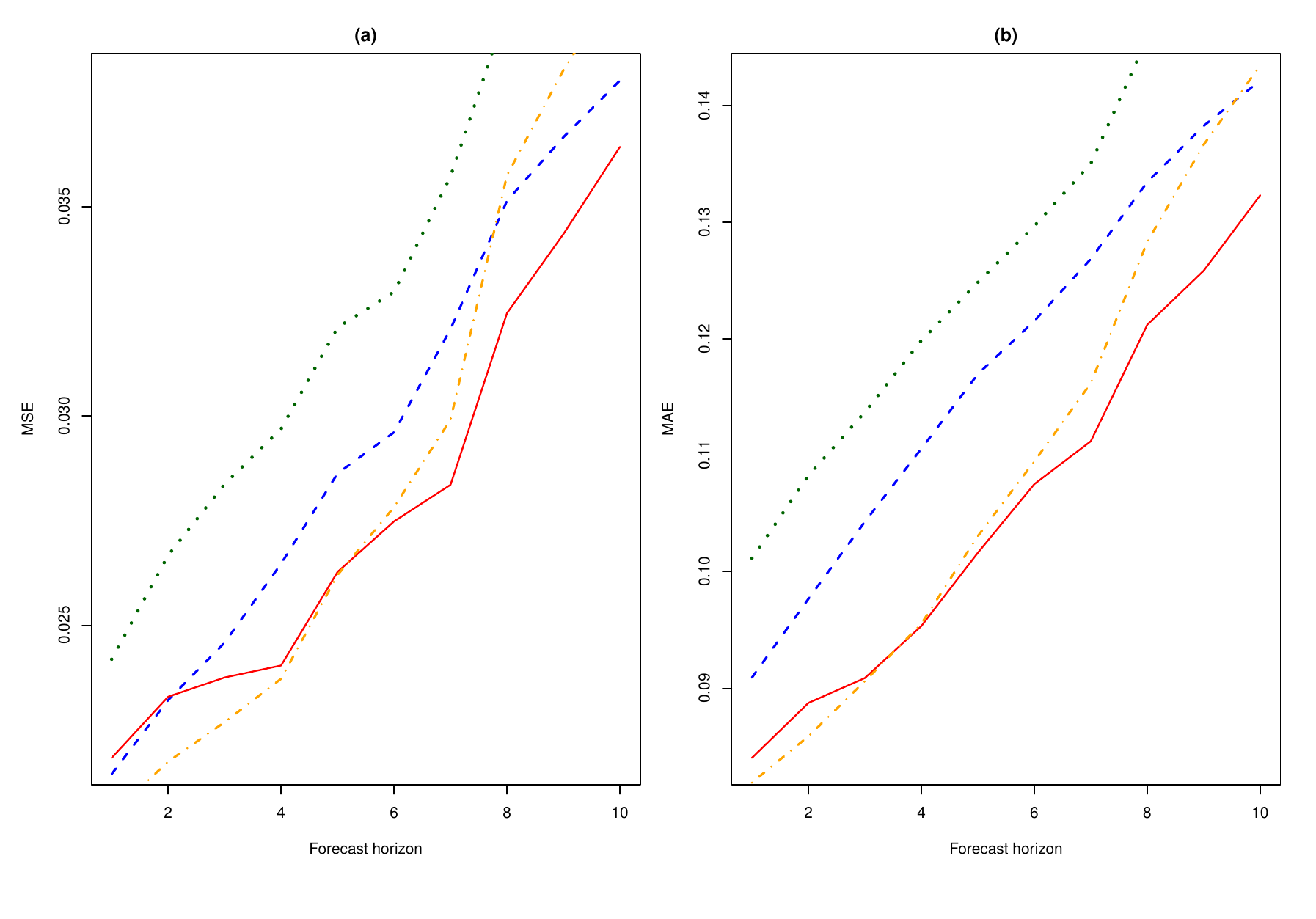}
\caption{(a) MSE and (b) MAE plots for Belgian mortality of the HUts (red solid line), HU (blue dashed line), HUrob (green dotted line), and the wHU (orange dash-dotted line) models}
\label{fig: bgmME}
\end{figure}

\begin{figure}[htbp!]
\centering
\includegraphics[width=0.75\linewidth]{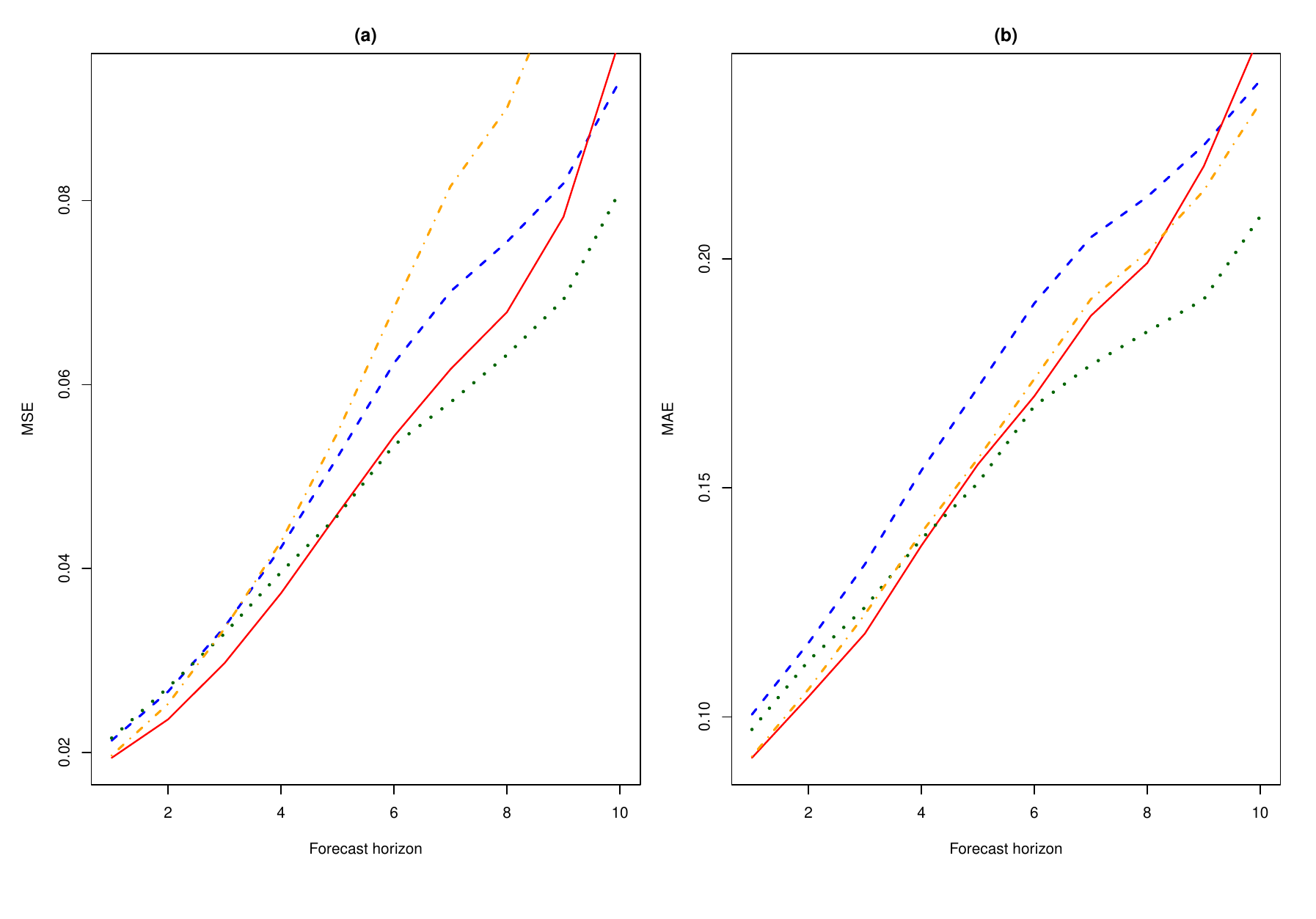}
\caption{(a) MSE and (b) MAE plots for Bulgarian mortality of the HUts (red solid line), HU (blue dashed line), HUrob (green dotted line), and the wHU (orange dash-dotted line) models}
\label{fig: bgrME}
\end{figure}

\begin{figure}[htbp!]
\centering
\includegraphics[width=0.75\linewidth]{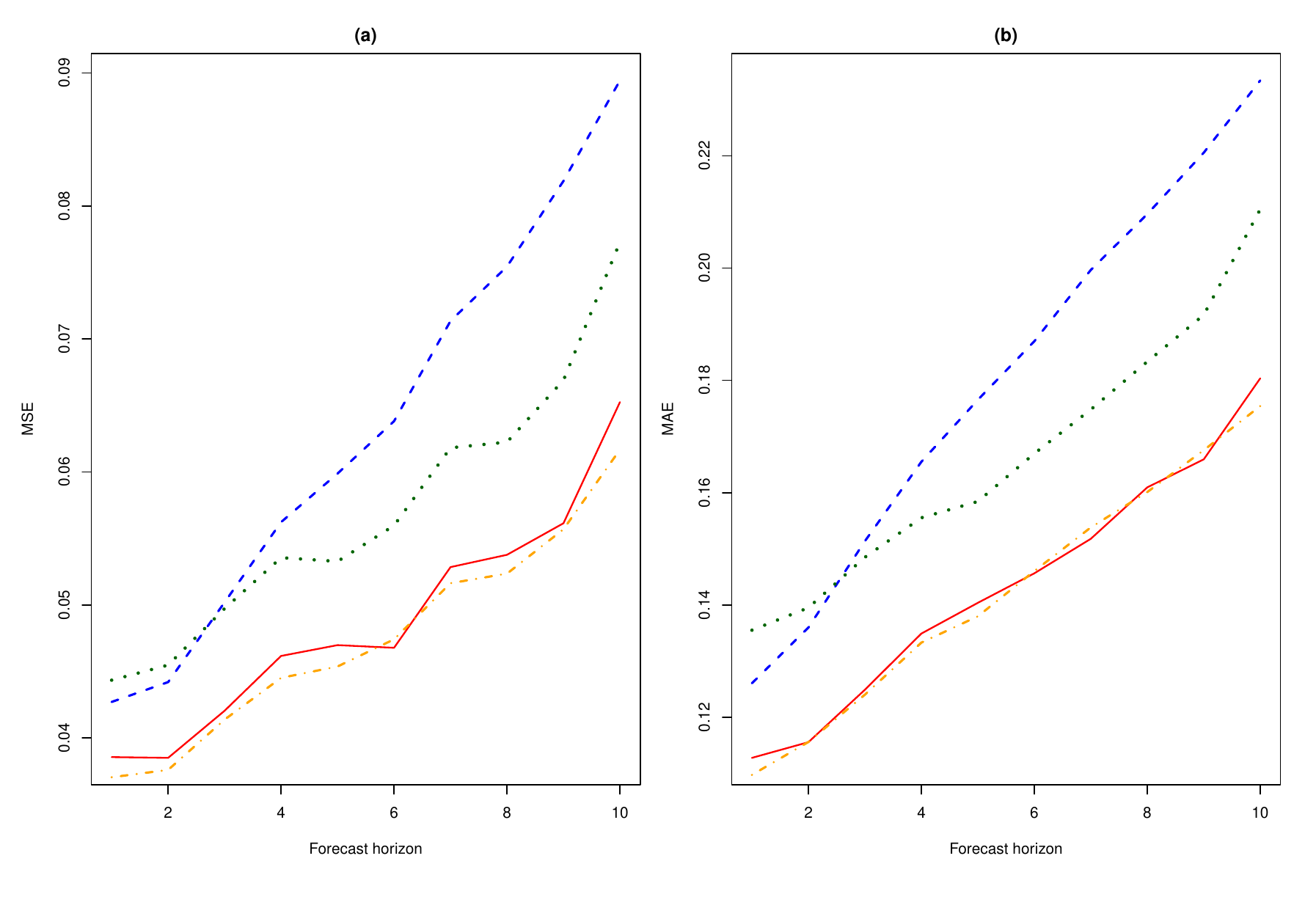}
\caption{(a) MSE and (b) MAE plots for Finnish mortality of the HUts (red solid line), HU (blue dashed line), HUrob (green dotted line), and the wHU (orange dash-dotted line) models}
\label{fig: finME}
\end{figure}

\begin{figure}[htbp!]
\centering
\includegraphics[width=0.75\linewidth]{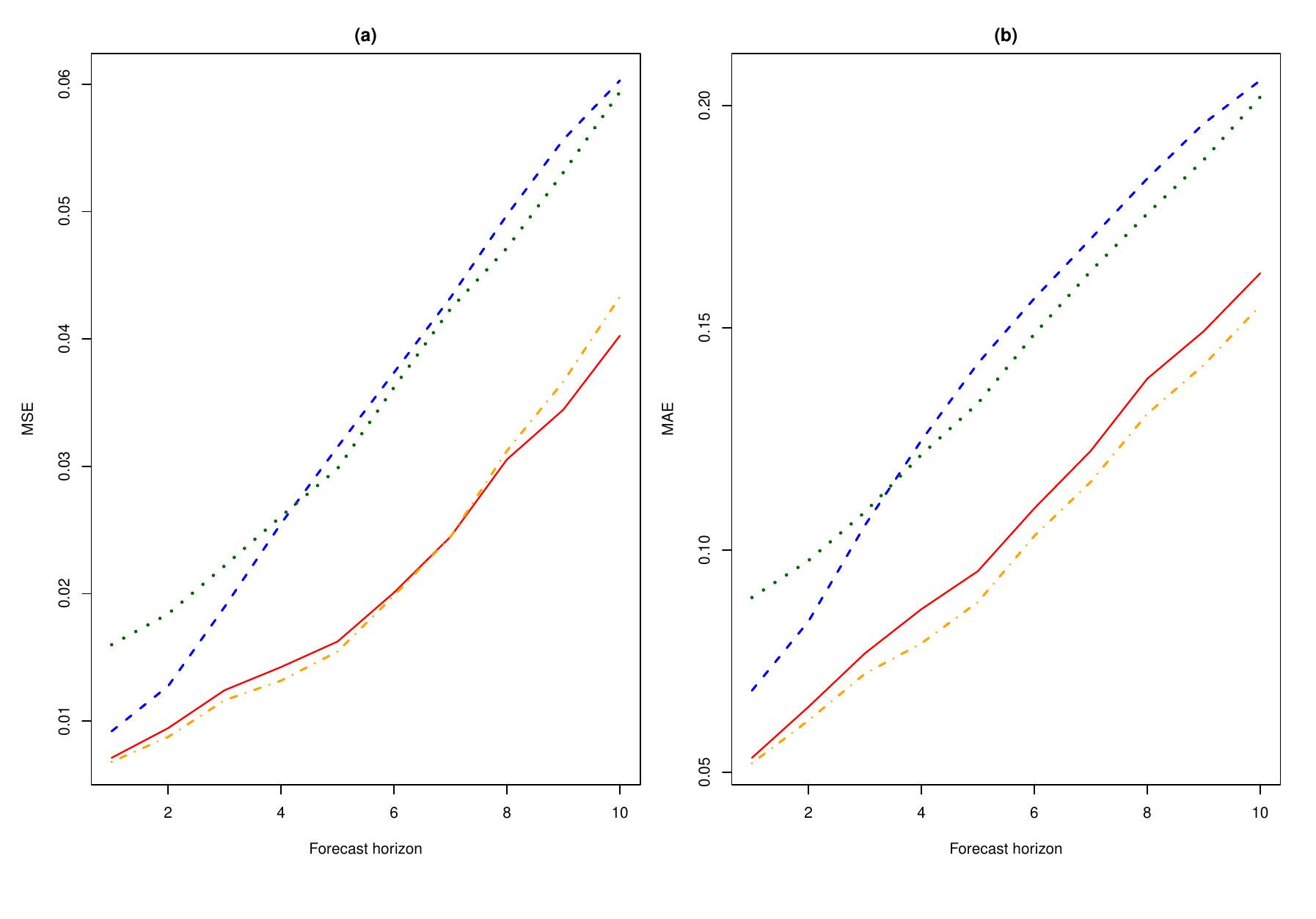}
\caption{(a) MSE and (b) MAE plots for Italian mortality of the HUts (red solid line), HU (blue dashed line), HUrob (green dotted line), and the wHU (orange dash-dotted line) models}
\label{fig: ityME}
\end{figure}

\begin{figure}[htbp!]
\centering
\includegraphics[width=0.75\linewidth]{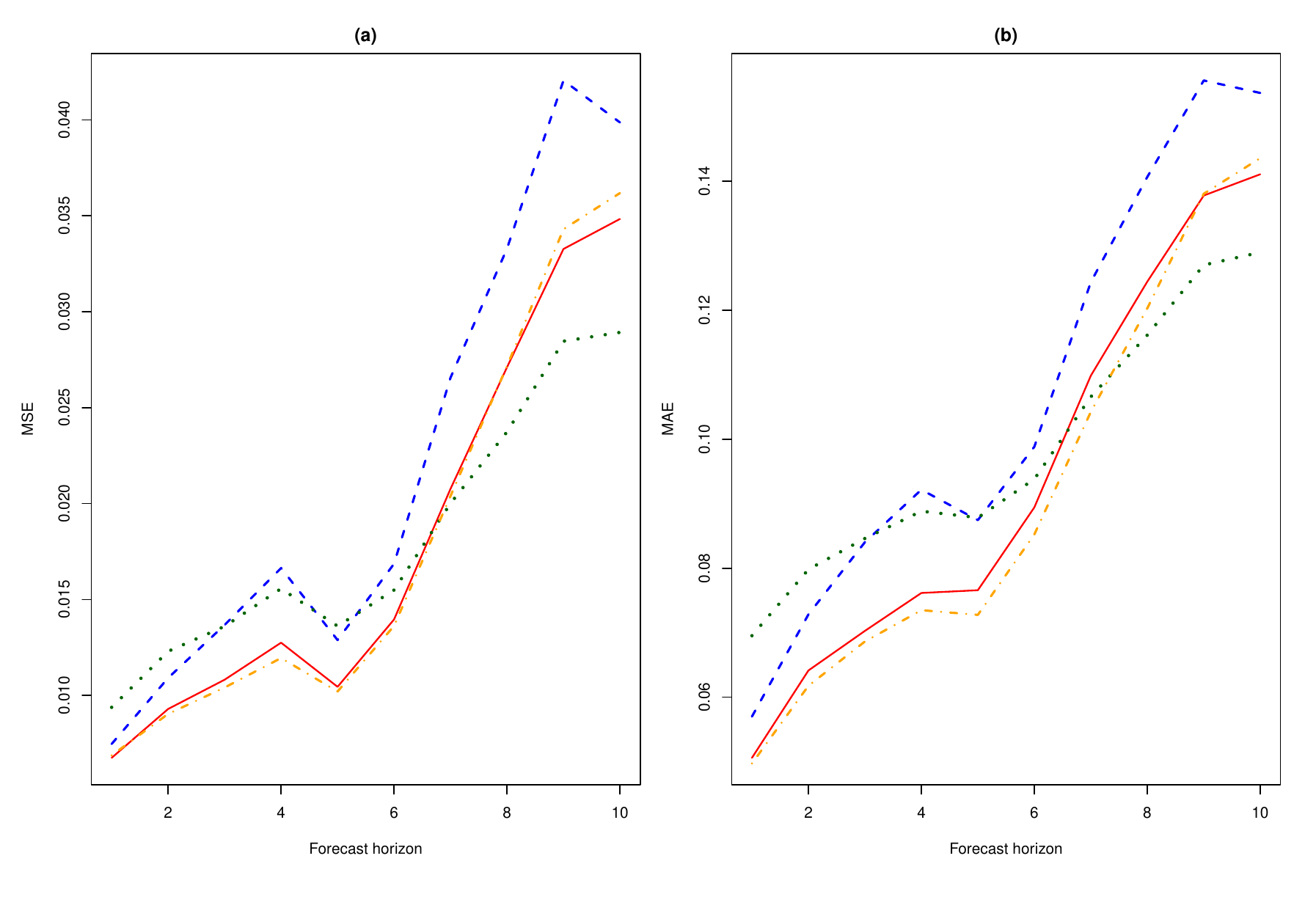}
\caption{(a) MSE and (b) MAE plots for Japanese mortality of the HUts (red solid line), HU (blue dashed line), HUrob (green dotted line), and the wHU (orange dash-dotted line) models}
\label{fig: jpME}
\end{figure}

\begin{figure}[htbp!]
\centering
\includegraphics[width=0.75\linewidth]{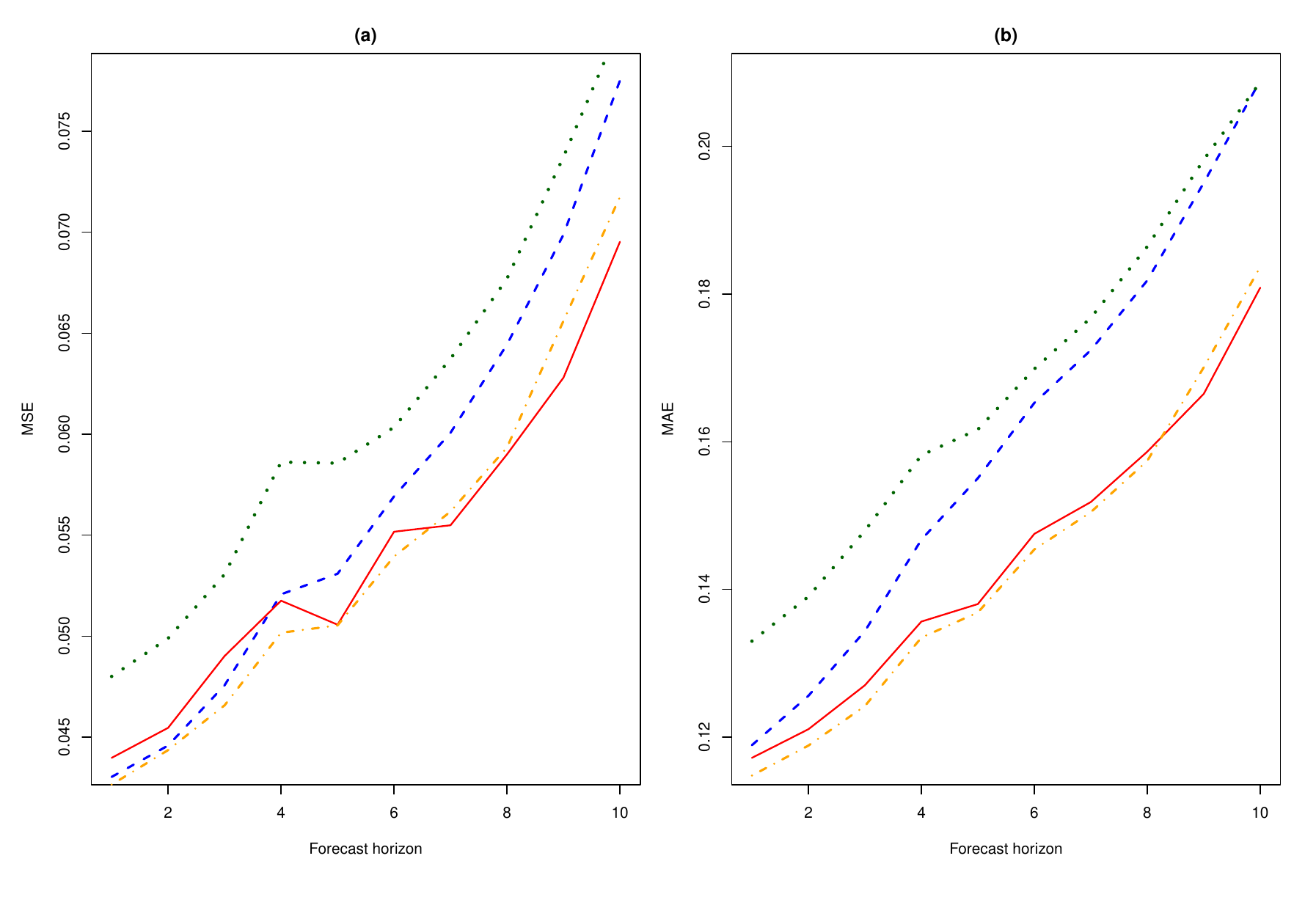}
\caption{(a) MSE and (b) MAE plots for Norwegian mortality of the HUts (red solid line), HU (blue dashed line), HUrob (green dotted line), and the wHU (orange dash-dotted line) models}
\label{fig: norwayME}
\end{figure}
\FloatBarrier

\subsection{Mean error plots}\label{AppendixF}
The mean error (ME) metric can be used to identify whether the models tend to overpredict or underpredict mortality rates at specific ages. It is computed with:
\begin{equation}
    ME(h,x_i) = \frac{1}{q}\sum_{t=1}^{q} (y_t(x_i)-\hat{y}_{t|t-h}(x_i)). \nonumber
\end{equation}
\begin{figure}[htbp]
    \centering
    \includegraphics[width=0.75\linewidth]{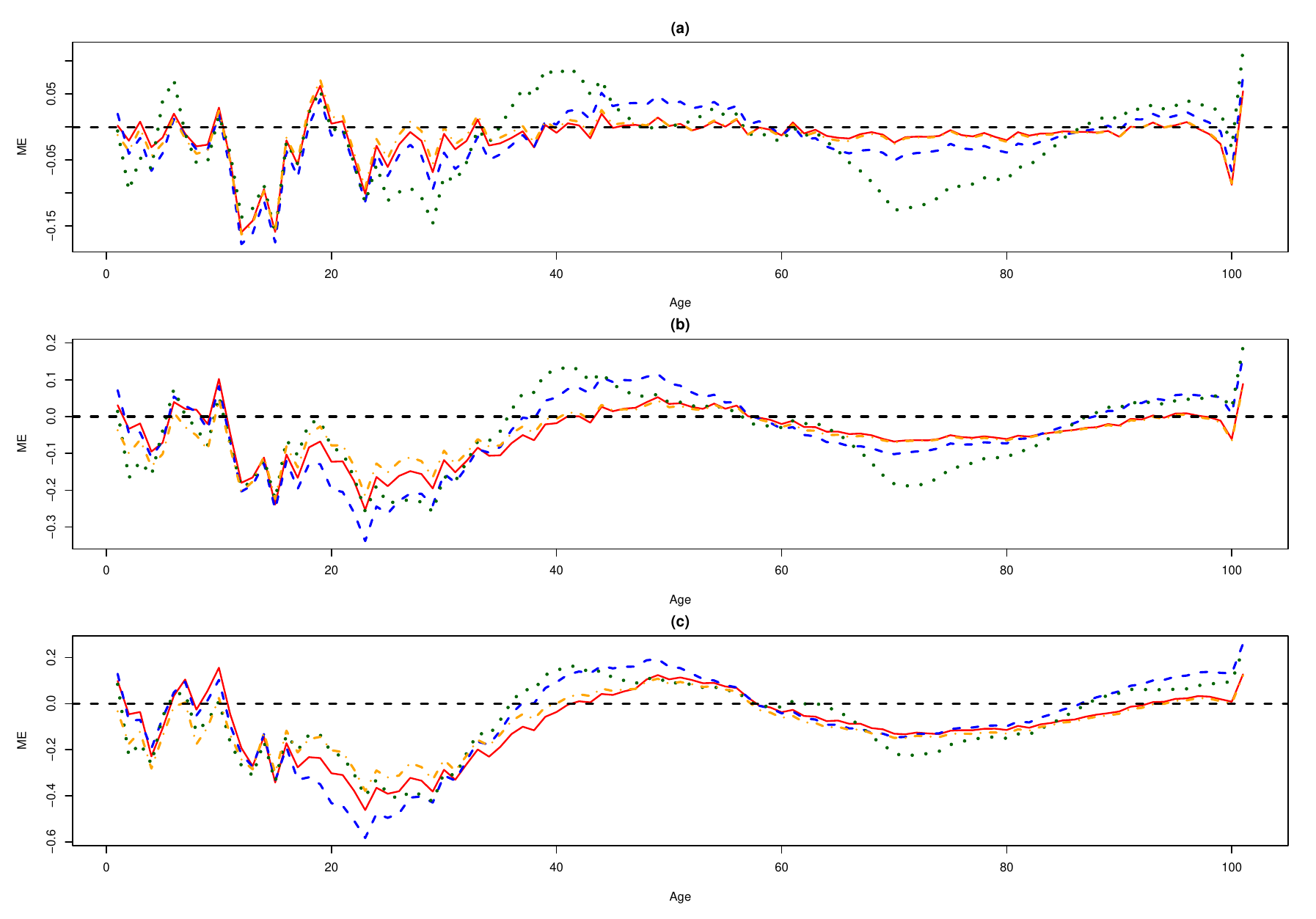}
    \caption{Mean error plot of Australian mortality at (a) h=1; (b) h=5; (c) h=10}
    \label{fig:ausme}
\end{figure}

\begin{figure}[htbp]
    \centering
    \includegraphics[width=0.75\linewidth]{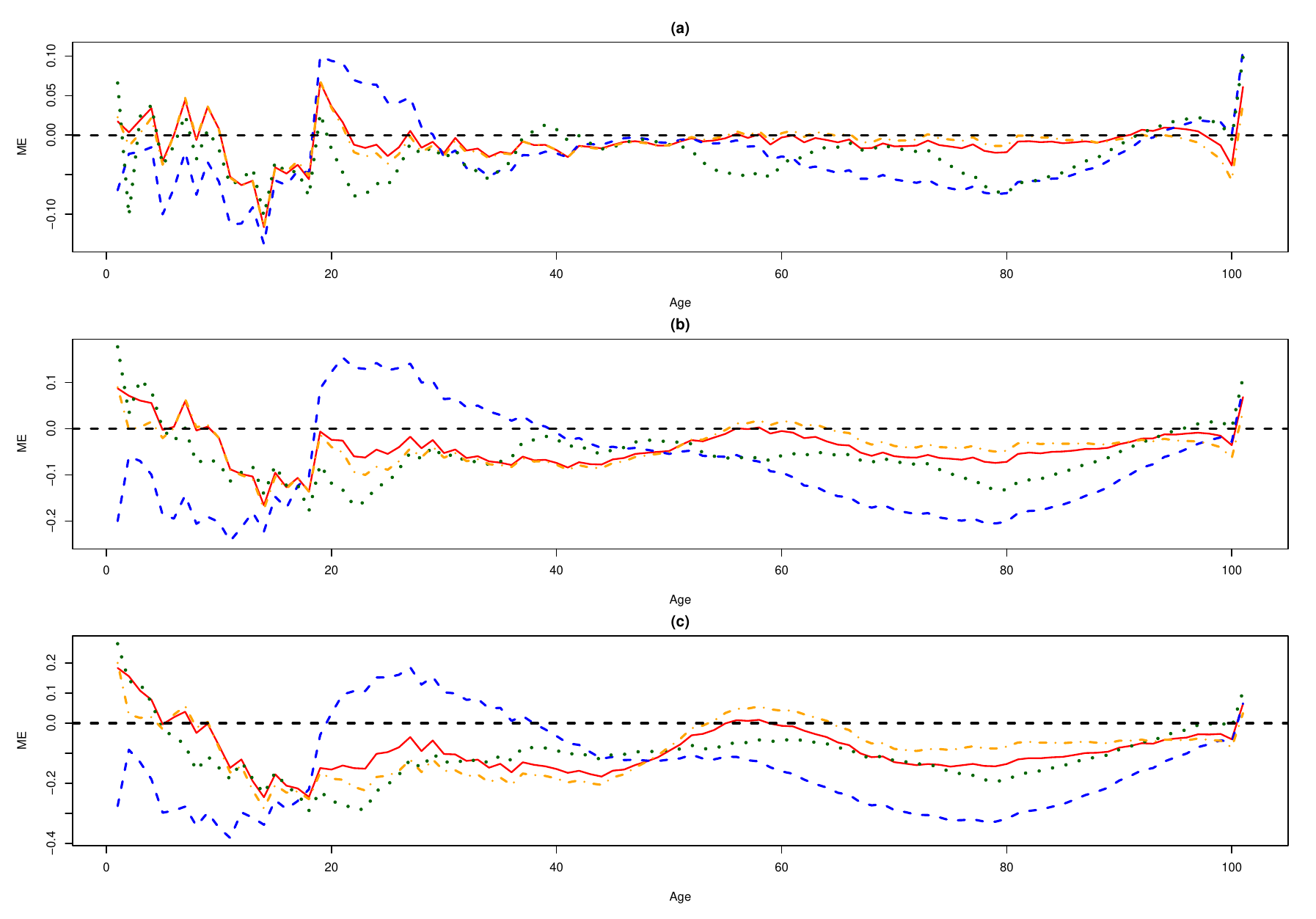}
    \caption{Mean error plot of French mortality at (a) h=1; (b) h=5; (c) h=10}
    \label{fig:frme}
\end{figure}

Upon a closer inspection of the ME plots across different age groups of the French mortality data in Figure \ref{fig:frme}, the HUts model generally demonstrates minimal bias across ages. For one-step-ahead forecasts (Figure \ref{fig:frme} (a)), the ME plot indicates that the HUts model’s forecasts are mostly unbiased, with slight dips around age 15 and peaks near ages 20 and 100. The wHU model shows a similar pattern, while the HU and HUrob models generally underpredict mortality across ages.

For five-step-ahead forecasts (Figure \ref{fig:frme} (b)), the ME plot for the HUts model remains close to zero, indicating sustained minimal bias. The wHU model continues to exhibit similar trends, whereas the HU and HUrob models persist in underpredicting mortality. In the ten-step-ahead forecasts (Figure \ref{fig:frme} (c)), the HUts model’s ME plot shows a slight bias at extreme ages but maintains near-zero mean error across most ages, suggesting accurate and unbiased forecasts overall. The HU and HUrob models display larger biases, contributing to less reliable predictions.

Overall, these observations suggest that the HUts model produces mostly unbiased forecasts, especially as the forecast horizon increases. While the wHU model also performs well at shorter horizons, the HUts model becomes more favorable for longer-term predictions due to its minimal bias across ages.

Similarly, the ME plots for the Australian mortality data in Figure \ref{fig:ausme} highlight the effectiveness of the HUts model, particularly in short-term forecasting. For one-step-ahead forecasts (Figure \ref{fig:ausme} (a)), the ME plot reveals minimal bias across ages for the HUts model, mirroring the performance of the wHU model. In contrast, the HU and HUrob models exhibit a tendency to underpredict mortality.

At the five-step-ahead horizon (Figure \ref{fig:ausme} (b)), the ME plot for the HUts model indicates underprediction biases in the 20-40 age range, contributing to increased errors in this group. By the ten-step-ahead forecasts (Figure \ref{fig:ausme} (c)), this underprediction becomes more pronounced in the same age range, significantly affecting the model’s overall accuracy. Despite this, the HUts model maintains reasonable performance across other age groups, though the wHU model generally achieves better results at longer horizons.

\FloatBarrier
\subsection{The Hyndman-Ullah with randomised signatures model}\label{AppendixG}
Randomised signatures are a computationally efficient alternative to the truncated signatures from rough path theory. Randomized signatures projects the high-dimensional signature space into a lower-dimensional space using random linear projections. These projections approximate the original signature’s properties and expressiveness, potentially reducing the computational need. This is achieved by evolving a system of random differential equations driven by the input path, resulting in a feature representation that maintains the theoretical guarantees of the signature while being computationally tractable for downstream tasks such as classification or system identification. See \citet{CompagnoniRSL,cuchiero2024jointcalibrationspxvix,cuchiero2024signaturemethodsstochasticportfolio,Cuchiero2020DiscreteTimeSA} for a more detailed introduction on randomised signatures.
The following is an algorithm to generate randomised signatures:

\begin{algorithm}
\caption{Generate randomized signature \citep{CompagnoniRSL}}
\label{alg:randomized_signature}
\begin{algorithmic}[1]
\REQUIRE $X \in \mathbb{R}^d$ sampled at $0 = t_0 < t_1 < \dots < t_N = T$, randomised signature dimension $k$, activation function $\sigma$.
\STATE Initialize $Z_{t_0} \in \mathbb{R}^k$, $A_i \in \mathbb{R}^{k \times k}$, $b_i \in \mathbb{R}^k$ to have i.i.d. standard normal entries for $i \in \{1, \dots, d\}$.
\FOR{$n = 1$ to $N$}
    \STATE Compute:
    \[
    Z_{t_n} = Z_{t_{n-1}} + \sum_{i=1}^d \sigma(A_i Z_{t_{n-1}} + b_i) (X^i_{t_n} - X^i_{t_{n-1}})
    \]
\ENDFOR
\end{algorithmic}
\end{algorithm}

By replacing the usage of truncated signatures with randomised signature, we can model the morality rates using the Hyndman-Ullah with randomised signatures (HUrs) model. We set the number of signature dimension $k=75$, and use the linear activation function. The MSE results are tabulated below.

\begin{table}[htbp!]
\centering
\caption{MSE: HUts, HUrs, and wHU across different countries for $h=1,5,10$}
\label{tab: MSEHURs}
\begin{tabular}{lccccccccc}
\toprule
\multirow{2}{*}{Country} & \multicolumn{3}{c}{h=1} & \multicolumn{3}{c}{h=5} & \multicolumn{3}{c}{h=10} \\ \cmidrule(lr){2-4} \cmidrule(lr){5-7} \cmidrule(lr){8-10}
& HUts & HUrs & wHU & HUts & HUrs & wHU & HUts & HUrs & wHU \\
\midrule
Australia       & 0.00982 & 0.01062 & 0.00952 & 0.01835 & 0.03038 & 0.01642 & 0.0418 & 0.08545 & 0.0351 \\
Belgium         & 0.02183 & 0.02371 & 0.02047 & 0.02627 & 0.04172 & 0.02620 & 0.0364 & 0.09131 & 0.0404 \\
Bulgaria        & 0.01942 & 0.01930 & 0.01968 & 0.04595 & 0.04442 & 0.05478 & 0.0976 & 0.08949 & 0.1280 \\
Denmark         & 0.03636 & 0.03914 & 0.03590 & 0.04891 & 0.07024 & 0.05007 & 0.0764 & 0.14972 & 0.0825 \\
Finland         & 0.03856 & 0.04569 & 0.03705 & 0.04698 & 0.07012 & 0.04538 & 0.0653 & 0.12988 & 0.0617 \\
France          & 0.00398 & 0.00615 & 0.00399 & 0.00882 & 0.02614 & 0.00978 & 0.0201 & 0.08445 & 0.0242 \\
Ireland         & 0.05062 & 0.05275 & 0.05161 & 0.07035 & 0.09055 & 0.07785 & 0.1111 & 0.17584 & 0.1413 \\
Italy           & 0.00711 & 0.00995 & 0.00680 & 0.01625 & 0.03723 & 0.01546 & 0.0403 & 0.11560 & 0.0433 \\
Japan           & 0.00674 & 0.00680 & 0.00687 & 0.01045 & 0.01000 & 0.01020 & 0.0348 & 0.03289 & 0.0362 \\
Netherlands     & 0.01242 & 0.01435 & 0.01143 & 0.01785 & 0.03353 & 0.01859 & 0.0326 & 0.09922 & 0.0377 \\
Norway          & 0.04397 & 0.04692 & 0.04264 & 0.05056 & 0.06853 & 0.05053 & 0.0695 & 0.13147 & 0.0718 \\
United States   & 0.00133 & 0.00162 & 0.00129 & 0.00746 & 0.01023 & 0.00896 & 0.0226 & 0.03197 & 0.0305 \\
\midrule
Mean            & 0.02101 & 0.02308 & 0.02061 & 0.03068 & 0.04443 & 0.03202 & 0.0540 & 0.1014  & 0.0611 \\
\bottomrule
\end{tabular}
\end{table}
As observed in Table \ref{tab: MSEHURs}, only Bulgaria appears to have experienced an improvement in forecasting accuracy. Nevertheless, the randomized signature requires numerous variables that require predetermination, necessitating further investigation to determine the optimal set of hyperparameters.

\end{appendix}

\end{document}